\documentclass [12pt]{iopart}
\usepackage {iopams}
\usepackage{setstack}
\usepackage{amstext}
\usepackage{dsfont}
\usepackage{amsfonts}
\usepackage{graphicx}
\usepackage{subfigure}
\usepackage{array}
\usepackage{appendix}
\usepackage{mathrsfs}
\usepackage{color}
\usepackage{adjustbox,lipsum}
\usepackage{upgreek}
\usepackage{algorithm}
\usepackage{algpseudocode}
\usepackage{makecell}
\usepackage{bm}

\usepackage{booktabs}

\usepackage{amsthm}
\usepackage{amssymb}
\usepackage{latexsym}
\usepackage{hyperref}
\usepackage{float}
\usepackage{verbatim}
\usepackage{bbold}
\usepackage[normalem]{ulem}

\usepackage{upgreek}

\newcommand{\bra}[1]{\left<#1\right|}
\newcommand{\ket}[1]{\left|#1\right>}
\newcommand{\abs}[1]{\bigl|#1\bigr|}
\newcommand{\norm}[1]{\left\| #1 \right\|}

\newcommand{\braket}[2]{\left<{#1}|{#2}\right>}
\newcommand{\ketbra}[2]{\ket{#1}\!\!\bra{#2}}

\newtheorem{theorem}{Theorem}
\newtheorem{proposition}{Proposition}
\newtheorem{lemma}{Lemma}
\newtheorem{corollary}{Corollary}

\begin{document}
\title[]{Causal Consistency and Cost of Quantum Control}

\author{Jeongho~Bang$^{1,2}$, Kyoungho~Cho$^{1,3}$, and Jeongwoo~Jae$^{4}$}
\address{$^1$ Institute for Convergence Research and Education in Advanced Technology, Yonsei University, Seoul 03722, Republic of Korea}
\address{$^2$ Department of Quantum Information, Yonsei University, Incheon 21983, Republic of Korea}
\address{$^3$ Department of Statistics and Data Science, Yonsei University, Seoul 03722, Republic of Korea}
\address{$^4$R\&D Center, Samsung SDS, Seoul 05510, Republic of Korea}

\vspace{10pt}

\begin{indented}
\item {\em Correspondence and requests for materials should be addressed to J.B, K.C., and J.J.}
\end{indented}

\ead{\mailto{jbang@yonsei.ac.kr}, \mailto{khcho23@yonsei.ac.kr}, and \mailto{jeongwoo.jae@samsung.com}}

\begin{abstract}
An unknown quantum state can supply the information needed to control a computation, but how much quantum data must be consumed to perform a prescribed action? We connect this resource question to causal consistency within the quantum state space. No-signaling requires identical observable predictions for remotely chosen ensemble preparations of the same complete input; quantum distinguishability then converts this requirement into finite-copy bounds. For amplitude amplification, we identify an exact low-copy obstruction and establish a dimension-independent, optimal quadratic dependence of the copy cost on angular gain at fixed accuracy, uniformly over arbitrary pure inputs. A coherent protocol attains this scaling without reconstructing a classical state description. Prior knowledge of a real two-level family additionally permits exponentially decreasing error with the copy budget at fixed gain. These results separate the resources needed to perform a specified quantum operation from those needed for universal state reconstruction. The central connection is operational: physical consistency restricts which distinctions a controller can expose, and the information carried by its complete input determines the cost of doing so. This links a causal requirement to quantitative limits on converting quantum data into computational control.
\end{abstract}

\maketitle

\section{Introduction}\label{Sec:Intro}

Quantum control is usually specified by a known operation and an input on which it acts. A different resource problem arises when the operation itself depends on an unknown quantum state. Such a state can define a reference for a reflection, an intermediate step of a computation, or a target for an adaptive controller. Describing the desired action is straightforward; supplying the information needed to enact it is not. How many copies of an unknown state are required before it can serve as a control reference? This question places the physical representation of information at the interface between quantum statistics and computation.

State reconstruction and a prescribed control action impose different output requirements. Tomography returns a classical description that supports independently chosen subsequent operations. A controller may instead consume copies of the state as a quantum program for one specified action. Both tasks can be defined over all pure inputs, yet universality of the input promise does not make their information requirements identical. We therefore count the complete quantum input supplied to a controller and distinguish protocols that reconstruct a state from those that directly use it. This makes the resource cost a property of an operational task rather than of an unspecified ability to learn.

A physical consistency condition links these tasks to causality. Measurements on a distant purification can prepare different ensemble decompositions of the same local density operator~\cite{hughston1993complete}. A local device cannot reveal which decomposition was chosen unless the remote outcome is communicated. This no-signaling requirement underlies restrictions on nonlinear dynamics~\cite{gisin1990weinberg,simon2001nosignaling}, state estimation~\cite{han2010optimal}, and discrimination~\cite{bae2011nosignaling}. Bao, Bouland, and Jordan connected search speedups and signaling for specified modifications of quantum mechanics~\cite{bao2016grover}. When a controller receives additional state copies, however, causal consistency must be tested on that complete input. The distinguishability of the supplied quantum data is then the relevant statistical resource.

Reflections about unknown pure states provide a concrete setting for this connection. Every fixed reflection is a legitimate unitary, whereas a universal processor cannot implement the corresponding unknown-state reflection exactly from a finite quantum program~\cite{PhysRevLett.79.321,kumar2011quantum}. In amplitude amplification, replacing the fixed reflection by one about the preceding output formally triples the amplification angle at each step. Estimation-based proposals make the required intermediate control information explicit~\cite{Baek2025asd}. The number of formal updates nevertheless leaves open the physical cost of acquiring, preparing, and using that information.

Here, we show that a controller must receive enough copies of an unknown pure state to distinguish inputs whose prescribed outputs are easy to tell apart. For amplitude amplification, we construct two preparations that give identical measurement statistics with too few copies, even though ideal amplification would yield perfectly distinguishable outputs. For arbitrary pure inputs at fixed accuracy, the minimum copy count grows as the square of the factor by which the angle is amplified, independently of the dimension of the state space. We give a protocol that attains this scaling by using the copies directly. For real superpositions of two known basis states, as for a photon with unknown linear polarization, we also determine the cost of higher accuracy. At fixed amplification, the optimal copy count for this family grows logarithmically with the inverse of the allowed error. Finally, we examine how estimation errors accumulate over successive control steps and account for the cost of preparing the states used in those steps.

The physical principle, the statistical bound, and the implementation have distinct roles. No-signaling supplies ensemble consistency within the quantum state space, while quantum distinguishability determines its finite-copy scale. Density matrix exponentiation provides a constructive upper bound~\cite{lloyd2014quantum,kimmel2017hamiltonian,go2025sample}; sample-to-query lifting and pure-state samplization relate copy access to oracle access~\cite{wang2025lifting,wang2026samplizer}. Density matrix exponentiation and pure-state samplization provide channel simulations acting on arbitrary data. Sample-to-query lifting instead relates sample lower bounds to query lower bounds under specified oracle access. Our amplification task requests one prescribed output from copies of the same unknown state. A lower bound for the stronger channel task therefore does not by itself establish the copy cost of this output task. Our task-specific results are the exact finite-copy ensemble analysis, the matching fixed-error bound for the prescribed amplification output, and the precision-optimal real-circle construction together with an obstruction to exact universal heralding. The general expansion theorem expresses their common necessary condition. The resulting connection between physics and computation is a quantitative restriction on the conversion of input distinguishability into control: a desired action must be supported by sufficient information in the physical systems that implement it.

\section{State-dependent control and the sample model}\label{sec:model}

Let $\mathcal H$ have dimension $N\ge2$, let $\hat{\Pi}$ be a fixed nontrivial projector, and define $\hat{R}_{\Pi}=\hat{\mathds{1}}-2\hat{\Pi}$. Every pure input with nonzero components in both subspaces can be written as
\begin{eqnarray}
\ket{\psi} = \sin\theta\ket{g_\psi} + \cos\theta\ket{b_\psi}, \quad (0<\theta<\pi/2),
\label{eq:input_decomposition}
\end{eqnarray}
where $\ket{g_\psi}$ and $\ket{b_\psi}$ are its normalized projections. For an odd integer $k=2q+1$, define the target pure-state transformation by
\begin{eqnarray}
\ket{A_k(\psi)} = \sin(k\theta)\ket{g_\psi} + \cos(k\theta)\ket{b_\psi}.
\label{eq:amplification_target}
\end{eqnarray}
The target density operator is well defined at the endpoints by continuity. These definitions cover all pure inputs; the component directions need not be known. Standard reflection geometry~\cite{grover1996fast,brassard2000quantum} gives
\begin{eqnarray}
\ket{A_k(\psi)} = (-\hat{R}_\psi\hat{R}_{\Pi})^q\ket{\psi}, \quad \hat{R}_\psi=\hat{\mathds{1}}-2\ketbra{\psi}{\psi}.
\label{eq:fixed_reflection_representation}
\end{eqnarray}
For completeness, in the ordered basis $\{\ket{b_\psi},\ket{g_\psi}\}$ the two-reflection operator is
\begin{eqnarray}
-\hat{R}_\psi \hat{R}_\Pi =  \left( \begin{array}{cc} \cos2\theta & -\sin2\theta \\ \sin2\theta & \cos2\theta \end{array} \right).
\label{eq:rotation_matrix}
\end{eqnarray}
It rotates the vector $(\cos\theta,\sin\theta)^T$ to angle $(2q+1)\theta$ after $q$ applications, proving Eq.~(\ref{eq:fixed_reflection_representation}). The previous-output update $\ket{\psi_{r+1}}=-\hat{R}_{\psi_r}\hat{R}_{\Pi}\ket{\psi_r}$ instead rotates a signed angle $\vartheta$ to $3\vartheta$ in one step. Repeating it gives $k=3^R$. Signed angles retain this identity after an overshoot; ``amplification'' refers to angular gain and does not require the target population to increase for every input.

The directions $\ket{g_\psi}$ and $\ket{b_\psi}$ are arbitrary unknown vectors in the projector subspaces. The processor must therefore implement this geometry over the full pure-state space.

We use two levels of physical assumptions. The causal converse assumes the standard quantum state space, density operators as outputs, ordinary probabilistic averaging, and remote preparation of ensemble decompositions. A deterministic device has an output rule on the full state space of its $M$ input registers and obeys no-signaling. Its accuracy is required only when those registers contain identical copies of a promised pure state. The full domain is a consistency requirement, not an additional accuracy promise: auxiliary inputs may be processed arbitrarily, but must have normalized outputs. Linearity and complete positivity are not presupposed in this converse.

For achievability and the minimum quantum copy cost, we restrict to standard quantum realizations. Our universal benchmark supplies $M$ independent copies of the same initial state $\hat{\rho}_\psi=\ketbra{\psi}{\psi}$ to a deterministic completely positive trace-preserving (CPTP) processor $\mathcal L_M$, independent of $\psi$, satisfying
\begin{eqnarray}
\sup_{\psi}D\left(\mathcal L_M(\hat{\rho}_\psi^{\otimes M}), \hat{\rho}_{A_k(\psi)}\right) \le \eta,
\label{eq:implementation_goal}
\end{eqnarray}
where $D(\hat{\rho},\hat{\sigma})=\frac12\norm{\hat{\rho}-\hat{\sigma}}_1$. All copies, including an active input register, are counted. Ancillas and classical randomness carry no prior information about $\psi$; $\hat{R}_\Pi$ is available. Classical measurement, estimation, and feed-forward are included in this model, as is direct coherent processing of the copies. Producing a reusable classical state description is a stronger, separately specified output requirement.

The budget $M$ is a fixed maximum and the output is unconditional; postselected success defines a different task. Access to a preparation unitary and its inverse is also stronger than access to output copies. These specifications prevent preparation and postselection from supplying uncounted resources. \ref{app:setting} gives the operational details, the reflection geometry, and the finite-program obstruction to exact universal reflection.

\section{Finite-copy constraints from causal consistency}\label{sec:causality}

Let $T(\psi)$ be any prescribed quantum output state, with amplification corresponding to $T(\psi)=\hat\rho_{A_k(\psi)}$. For an ensemble $E_b=\{p_{b,j},\psi_{b,j}\}$, write its complete $M$-copy input and desired average output as
\begin{eqnarray}
\hat{\Omega}_b^{(M)}=\sum_j p_{b,j}\hat{\rho}_{\psi_{b,j}}^{\otimes M}, \quad \hat{\Gamma}_b=\sum_j p_{b,j}T(\psi_{b,j}).
\label{eq:ensembles}
\end{eqnarray}
If $\hat{\Omega}_0^{(M)}=\hat{\Omega}_1^{(M)}$, these are two decompositions of the same local state. Remote steering can select either decomposition without communicating its outcome. No-signaling therefore requires the same unconditioned local output. The same principle also constrains unequal inputs through a completion within the quantum state space.

\begin{lemma}[Ensemble consistency]\label{lem:ensemble_consistency}
Consider a deterministic device defined on the full quantum input space, with density operators as outputs and the ordinary probabilistic averaging rule for ensembles. If it obeys no-signaling under remote steering and approximates the target on every identical-copy input with error at most $\eta$, then
\begin{eqnarray}
D(\hat{\Gamma}_0,\hat{\Gamma}_1) \le D(\hat{\Omega}_0^{(M)},\hat{\Omega}_1^{(M)}) + 2\eta.
\label{eq:ensemble_bound}
\end{eqnarray}
\end{lemma}

\begin{proof}---Let $\hat\Phi_b$ be the actual output averaged over $E_b$. Convexity of the trace norm gives $D(\hat\Phi_b,\hat\Gamma_b)\le\eta$. Set $B=D(\hat\Omega_0^{(M)},\hat\Omega_1^{(M)})$. If $B=0$, remote steering of their common purification realizes either ensemble~\cite{hughston1993complete}, so no-signaling requires $\hat\Phi_0=\hat\Phi_1$ and the triangle inequality proves the claim.

For $B>0$, the Jordan decomposition is $\hat\Omega_0^{(M)}-\hat\Omega_1^{(M)}=\hat\Delta_+-\hat\Delta_-$, with $\hat\Delta_\pm\ge0$, orthogonal supports, and $\tr\hat\Delta_\pm=B$. The operators $\hat\sigma_\pm=\hat\Delta_\pm/B$ are states, and
\begin{eqnarray}
\hat{\Xi}=\frac{\hat\Omega_0^{(M)}+B\hat{\sigma}_-}{1+B} = \frac{\hat\Omega_1^{(M)}+B\hat\sigma_+}{1+B}.
\label{eq:causal_completion}
\end{eqnarray}
Refine each $\hat\sigma_\pm$ into any pure-state ensemble, and denote its average device output by $\hat X_\pm$. Both sides of Eq.~(\ref{eq:causal_completion}) now specify pure-state ensembles of $\hat\Xi$, each remotely preparable from a common purification. No-signaling and ordinary averaging imply
\begin{eqnarray}
\hat\Phi_0 + B\hat X_- = \hat\Phi_1 + B \hat X_+.
\label{eq:completed_output_identity}
\end{eqnarray}
Since outputs are normalized states, $D(\hat X_+,\hat X_-)\le1$. Thus, $D(\hat\Phi_0,\hat\Phi_1)\le B$, and two applications of the triangle inequality give Eq.~(\ref{eq:ensemble_bound}).
\end{proof}

The auxiliary inputs need not be identical-copy states, so the device must be defined beyond its accuracy promise. CPTP contractivity gives the same inequality. The causal proof instead identifies the role of ensemble equivalence within the quantum state space; no-signaling alone is not assumed to fix distinguishability in a general probabilistic theory. Physically, the causal requirement prevents a controller from exposing a remote choice between equivalent preparations of its complete input. Quantum-state geometry then determines how much distinguishability unequal inputs provide. The finite-copy bound combines these two ingredients. Explicit steering measurements and an extension to unequal average errors are given in~\ref{app:ensembles}.

Choose orthonormal $\ket{b}$ and $\ket{g}$ in the two projector subspaces, and let $\ket{\psi_\vartheta}=\cos\vartheta\ket{b}+\sin\vartheta\ket{g}$. Consider the uniform ensembles
\begin{eqnarray}
E_0 &=& \left\{\psi_{j\pi/k}:0\le j<k\right\}, \nonumber \\
E_1 &=& \left\{\psi_{(j+1/2)\pi/k}:0\le j<k\right\}.
\label{eq:polygon_ensembles}
\end{eqnarray}
Their ideal outputs are respectively $\ketbra{b}{b}$ and $\ketbra{g}{g}$. The following result identifies the exact sample threshold below which the two input ensembles are identical.

\begin{theorem}[Indistinguishable-input threshold]\label{thm:causal_threshold}
For odd $k\ge3$, any deterministic implementation of Eq.~(\ref{eq:amplification_target}) compatible with no-signaling on the standard quantum state space has worst-case trace-distance error $\eta\ge1/2$ if $M<k$. For qubit inputs this error floor is tight, attained by the constant output $\hat{\mathds{1}}/2$.
\end{theorem}

\begin{proof}---Each matrix entry of $\hat\rho_{\psi_\vartheta}$ is a linear combination of $1$ and $e^{\pm2i\vartheta}$. Tensor multiplication therefore gives operators $\hat C_s$ independent of $\vartheta$ such that
\begin{eqnarray}
\hat\rho_{\psi_\vartheta}^{\otimes M} = \sum_{s=-M}^{M}e^{2is\vartheta}\hat C_s.
\label{eq:fourier_support}
\end{eqnarray}
For any integer $s$ and offset $\alpha$,
\begin{eqnarray}
\frac1k\sum_{j=0}^{k-1}e^{2is(j+\alpha)\pi/k} = 
\left\{
\begin{array}{ll}
e^{2\pi is\alpha/k}, & k \mid s, \\
0, & k \nmid s. 
\end{array}
\right.
\label{eq:polygon_filter}
\end{eqnarray}
When $M<k$, only $s=0$ survives for both $\alpha=0$ and $\alpha=1/2$. Hence, $\hat\Omega_0^{(M)}=\hat\Omega_1^{(M)}=\hat C_0$. The signed-angle reflection identity gives the target at angle $k\vartheta$; thus the first polygon maps to $\ketbra{b}{b}$ and the second to $\ketbra{g}{g}$. Their distance is one, so Lemma~\ref{lem:ensemble_consistency} yields $1\le2\eta$. Finally, for $N=2$, $\hat{\mathds{1}}/2-\hat\rho$ has eigenvalues $\pm1/2$ for every pure $\hat\rho$. Its trace distance from any target is $1/2$, and the constant-output channel attains the bound.
\end{proof}

For $k=3$ and $M=1$, the two ensembles are the shifted trines at angles $\{0,\pi/3,2\pi/3\}$ and $\{\pi/6,\pi/2,5\pi/6\}$. Both average to $\hat{\mathds{1}}/2$, whereas the ideal update makes their outputs orthogonal. Thus, the physical obstruction is already present for a single qubit and does not depend on a search-oracle embedding.

The same witness yields a computable constraint at every $M$.

\begin{proposition}[Exact ensemble distinguishability]\label{prop:finite_bias}
For the ensembles in Eq.~(\ref{eq:polygon_ensembles}), define
\begin{eqnarray}
p_a = 2^{-M} \!\!\!\! \sum_{0 \le l \le M \atop l \equiv a \,({\rm mod} \, 2k)} \!\! {M \choose l}, \quad B_{M,k} = 2\sum_{a=0}^{k-1}\sqrt{p_a p_{a+k}},
\label{eq:finite_copy_bias}
\end{eqnarray}
where an empty sum is zero. Then $D(\hat{\Omega}_0^{(M)},\hat{\Omega}_1^{(M)})=B_{M,k}$, and any implementation with uniform error $\eta$ satisfies
\begin{eqnarray}
\eta\ge\frac{1-B_{M,k}}{2}.
\label{eq:finite_copy_error}
\end{eqnarray}
\end{proposition}

\begin{proof}---Set $\ket{y_\pm}=(\ket{b}\pm i\ket g)/\sqrt2$, and let $\ket{l}$ denote the normalized symmetric tensor with $l$ factors $\ket{y_-}$ and $M-l$ factors $\ket{y_+}$. With $b_l=2^{-M}{M \choose l}$,
\begin{eqnarray}
\ket{\psi_\vartheta}^{\otimes M}
=e^{-iM\vartheta}\sum_{l=0}^{M}\sqrt{b_l}e^{2il\vartheta}\ket l.
\label{eq:symmetric_expansion}
\end{eqnarray}
Polygon averaging retains a matrix entry $(l,l')$ exactly when $k\mid(l-l')$. In the block $l\equiv a\ ({\rm mod}\ k)$, define the orthogonal vectors
\begin{eqnarray}
\ket{v_a}={\!\!\!\!\!\!\!\!}\sum_{l\equiv a\ ({\rm mod}\ 2k)}{\!\!\!\!\!\!\!\!}\sqrt{b_l}\ket{l}, \quad \ket{w_a}={\!\!\!\!\!\!\!\!}\sum_{l \equiv a+k \, ({\rm mod} \, 2k)}{\!\!\!\!\!\!\!\!}\sqrt{b_l}\ket{l},
\label{eq:residue_vectors}
\end{eqnarray}
with $0\le l\le M$. Their squared norms are $p_a$ and $p_{a+k}$. The two averages have blocks $\ketbra{v_a+w_a}{v_a+w_a}$ and $\ketbra{v_a-w_a}{v_a-w_a}$, since the offset polygon contributes the sign $(-1)^{(l-l')/k}$. Their difference is $2(\ketbra{v_a}{w_a}+\ketbra{w_a}{v_a})$. Its nonzero eigenvalues are $\pm2\sqrt{p_a p_{a+k}}$; if either vector vanishes, the block is zero. Distinct residue blocks are orthogonal, so half the sum of their trace norms equals $B_{M,k}$. The output targets are orthogonal, and Lemma~\ref{lem:ensemble_consistency} proves Eq.~(\ref{eq:finite_copy_error}).
\end{proof}

At the first nontrivial budget $M=k$, only the residue pair $l=0,k$ contributes, giving $B_{k,k}=2^{1-k}$ and $\eta\ge1/2-2^{-k}$. In particular, $k=M=3$ gives $\eta\ge3/8$. The new angular mode becomes available at $M=k$, but its distinguishability is still exponentially small in $k$. This separates the appearance of information from its reliable use. \ref{app:ensembles} gives a channel saturating this boundary on a known real great circle; Eq.~(\ref{eq:finite_copy_error}) is not asserted to be the optimal error curve on the full input space.

\section{Resolution limits and optimal copy cost}\label{sec:optimal}

The causal threshold identifies when the input source contains no information at all about a relevant ensemble choice. A two-state argument makes the asymptotic sample requirement explicit. Under the causal assumptions of Lemma~\ref{lem:ensemble_consistency}, any $M$-copy implementation of a target transformation $T$ with error $\eta$ satisfies
\begin{eqnarray}
D\left(T(\psi),T(\phi)\right)
\le\sqrt{1-\abs{\braket{\psi}{\phi}}^{2M}}+2\eta.
\label{eq:pair_bound}
\end{eqnarray}
This follows from Eq.~(\ref{eq:ensemble_bound}) with singleton input ensembles, or directly from CPTP contractivity and the triangle inequality.

For an intuitive amplification example, compare $\ket b$ with $\cos[\pi/(2k)]\ket b+\sin[\pi/(2k)]\ket g$. These inputs differ by an angle of order $1/k$, but their required outputs are the orthogonal states $\ket b$ and $\ket g$. Their squared overlap after supplying $M$ copies is $\cos^{2M}[\pi/(2k)]=\exp[-\Theta(M/k^2)]$. Thus, a fixed output error below $1/2$ requires enough copies to make this small input difference resolvable, giving the quadratic scale.

The pair constraint applies to any prescribed state-dependent output, including mixed outputs and outputs on a larger Hilbert space. Its content can be stated as a general restriction on the resolution required by a control task. Define the Fubini--Study distance by $d_{\rm FS}(\psi,\phi)=\arccos\abs{\braket{\psi}{\phi}}$.

\begin{theorem}[Finite-scale expansion bound]\label{thm:finite_scale}
Let $T$ assign a quantum output state to every pure input in a specified promise set. Under the full-domain, mixing, and no-signaling assumptions of Lemma~\ref{lem:ensemble_consistency}, an unconditional $M$-copy implementation with error at most $\eta$ satisfies
\begin{eqnarray}
M\ge\left\lceil\frac{\ln\{1-[\Delta-2\eta]^2\}}{\ln(\cos^2\delta)}\right\rceil,
\label{eq:finite_scale_exact}
\end{eqnarray}
whenever two promised inputs have $0<\delta=d_{\rm FS}(\psi,\phi)<\pi/2$ and target separation $\Delta=D(T(\psi),T(\phi))>2\eta$. The bound is infinite if $\Delta-2\eta=1$. In particular, suppose a family $T_L$ admits such a pair with $\delta_L\le a/L$ and $\Delta_L\ge d_0$, where $a>0$ and $d_0>0$ are independent of $L$. For fixed $0<\eta<d_0/2$ and $L\ge2a$,
\begin{eqnarray}
M\ge\frac{L^2}{2a^2}\ln\frac{1}{1-(d_0-2\eta)^2}.
\label{eq:finite_scale_quadratic}
\end{eqnarray}
Thus, $M=\Omega_\eta(L^2)$, without a dimension-dependent prefactor.
\end{theorem}

\begin{proof}---The proof of Lemma~\ref{lem:ensemble_consistency} uses no property of the target beyond its being a density operator. Apply it to singleton ensembles. The distance between the pure tensor-power inputs is $\sqrt{1-\cos^{2M}\delta}$, so
\begin{eqnarray}
\cos^{2M}\delta\le1-(\Delta-2\eta)^2.
\label{eq:finite_scale_overlap}
\end{eqnarray}
If the right side vanishes, the inequality is impossible at finite $M$. Otherwise, taking logarithms and dividing by $\ln(\cos^2\delta)<0$ gives Eq.~(\ref{eq:finite_scale_exact}). For $0\le x\le1/2$, $\sin x\le x$ and $-\ln(1-u)\le u/(1-u)$ imply
\begin{eqnarray}
-\ln(\cos^2x)\le\frac{x^2}{1-x^2}\le2x^2.
\label{eq:finite_scale_log_bound}
\end{eqnarray}
Hence, the positive denominator $-\ln(\cos^2\delta_L)$ is at most $2a^2/L^2$, while the numerator $-\ln[1-(\Delta_L-2\eta)^2]$ is at least $-\ln[1-(d_0-2\eta)^2]$. Their ratio proves Eq.~(\ref{eq:finite_scale_quadratic}).
\end{proof}

The finite separation $d_0>2\eta$ is essential: a large derivative over an arbitrarily small output range need not impose a fixed-error sample cost. \ref{app:finite_scale} gives an explicit counterexample. This formulation turns ensemble consistency into a copy requirement for state-dependent tasks beyond amplitude amplification. For example, the replication target $T_r(\psi)=\hat\rho_\psi^{\otimes r}$ resolves inputs separated by order $1/\sqrt r$ into outputs with constant separation. Applying the theorem with $L=\sqrt r$ gives $M=\Omega_\eta(r)$ at any fixed $\eta<1/2$; retaining $r$ input copies supplies the matching upper bound. \ref{app:finite_scale} proves this statement, including its constants. Amplification realizes the same statistical obstruction with resolution $1/k$, yielding its quadratic copy cost.

\begin{theorem}[Optimal copy cost of amplification]\label{thm:optimal_copies}
Fix $0<\eta<1/2$. For every $N\ge2$ and nontrivial fixed $\hat{\Pi}$, the least initial-copy budget satisfying Eq.~(\ref{eq:implementation_goal}) obeys
\begin{eqnarray}
M_\star(k,\eta;N,\Pi)=\Theta_\eta(k^2)
\label{eq:optimal_scaling}
\end{eqnarray}
for odd $k\ge3$, with constants independent of $N$ and the rank of $\hat{\Pi}$. More explicitly,
\begin{eqnarray}
M_\star \ge \frac{\ln[4\eta(1-\eta)]}{\ln[\cos^2(\pi/(2k))]}, \quad M_\star \le 1+C\frac{(k-1)^2}{\eta}
\label{eq:explicit_bounds}
\end{eqnarray}
for a universal constant $C$. Optimality is asserted in $k$ at fixed error.
\end{theorem}

\begin{proof}---Choose normalized $\ket{b}\in\ker\hat{\Pi}$ and $\ket{g}\in\mathrm{ran}\hat{\Pi}$, and compare the inputs $\ket{b}$ and $\cos(\pi/(2k))\ket{b}+\sin(\pi/(2k))\ket{g}$. Their targets are $\ketbra{b}{b}$ and $\ketbra{g}{g}$. Eq.~(\ref{eq:pair_bound}) therefore implies
\begin{eqnarray}
1-2\eta \le \sqrt{1-\cos^{2M}(\pi/(2k))}, \quad \cos^{2M}(\pi/(2k)) \le 4\eta(1-\eta).
\label{eq:optimal_lower_proof}
\end{eqnarray}
Taking logarithms yields the stated lower bound, since $\ln[\cos^2(\pi/(2k))]<0$. Its large-$k$ value is
\begin{eqnarray}
\frac{4k^2}{\pi^2}\ln\frac{1}{4\eta(1-\eta)}[1+O(k^{-2})],
\label{eq:optimal_lower_expansion}
\end{eqnarray}
which is $\Omega_\eta(k^2)$ for every fixed $0<\eta<1/2$.

For achievability, retain one initial copy as data. Each approximate reflection consumes fresh copies of the same initial state as program registers; ancillas independent of $\psi$ are free in the sample count. Let $\hat{S}$ exchange the data and one program register. Applying $e^{-i\delta\hat{S}}$ and discarding the program induces the channel
\begin{eqnarray}
\mathcal{E}_{\delta,\rho}(\hat{X}) = \cos^2\delta\,\hat{X} + \sin^2\delta\,\hat{\rho}\,\tr\hat{X} - i\sin\delta\cos\delta\,[\hat{\rho},\hat{X}],
\label{eq:main_partial_swap}
\end{eqnarray}
where $\hat{\rho}=\hat{\rho}_\psi$. This formula follows by expanding $e^{-i\delta\hat{S}}=\cos\delta\,\hat{\mathds{1}}-i\sin\delta\,\hat{S}$ and tracing the program. Compare it with $\mathcal{U}_{\delta,\rho}(\hat{X})=e^{-i\delta\hat{\rho}}\hat{X}e^{i\delta\hat{\rho}}$.

We use the diamond norm: the induced trace norm of $\mathcal T\otimes\mathrm{id}_A$, maximized over arbitrary reference registers $A$. For a Hamiltonian of operator norm at most one, its commutator generator satisfies $\norm{-i[\hat{H},\cdot]}_\diamond \le 2$ by trace-norm submultiplicativity. The second derivative of its unitary channel consequently has diamond norm at most four: it is the generator squared composed with a CPTP channel, whose diamond norm is one. Integrating the second derivative bounds its first-order Taylor remainder by $2\delta^2$. Appending a program state and tracing that register preserve this bound, since both operations are CPTP. This applies to the partial-swap channel with $\norm{\hat{S}}_\infty=1$ and the ideal channel with $\norm{\hat{\rho}}_\infty=1$. Their zeroth-order terms coincide, and both first derivatives are $-i[\hat{\rho},\cdot]$. Therefore,
\begin{eqnarray}
\norm{\mathcal{E}_{\delta,\rho}-\mathcal{U}_{\delta,\rho}}_\diamond \le 4\delta^2.
\label{eq:main_partial_swap_bound}
\end{eqnarray}

Concatenate $m$ fresh-program steps with $\delta=\pi/m$. A telescoping expansion, using unit diamond norm for each channel, gives
\begin{eqnarray}
\norm{\mathcal{E}_{\pi/m,\rho}^{m}-\mathcal{U}_{\pi,\rho}}_\diamond\le\frac{4\pi^2}{m}.
\label{eq:main_reflection_bound}
\end{eqnarray}
Since $\hat{\rho}^2=\hat{\rho}$, $e^{-i\pi\hat{\rho}}=\hat{\mathds{1}}-2\hat{\rho}=\hat{R}_\psi$. Use this simulation for each of the $q=(k-1)/2$ reflections in Eq.~(\ref{eq:fixed_reflection_representation}), interleaved with the exact known reflection $\hat{R}_\Pi$. Another telescoping expansion bounds the total diamond error by $4\pi^2q/m$, hence the output trace-distance error by $2\pi^2q/m$. The diamond norm ensures these estimates remain valid with entangled intermediate registers.

Choosing
\begin{eqnarray}
m=\left\lceil\frac{2\pi^2q}{\eta}\right\rceil,
\quad
M=1+q\left\lceil\frac{2\pi^2q}{\eta}\right\rceil
\label{eq:main_constructive_copies}
\end{eqnarray}
therefore achieves error at most $\eta$ for every input. The first copy is the active data register, and all $qm$ program copies are consumed. Since $q\ge1$ and $\eta<1$, this count is at most $1+(2\pi^2+1)q^2/\eta$, proving the stated upper bound with $C=(2\pi^2+1)/4$. Every estimate is independent of the dimension and projector rank, completing the proof.
\end{proof}

Density matrix exponentiation~\cite{lloyd2014quantum,kimmel2017hamiltonian,go2025sample} and reflection samplization~\cite{wang2026samplizer} provide the simulation mechanism. Their channel guarantees protect arbitrary data, including a reference system. Eq.~(\ref{eq:implementation_goal}) asks for only one specified output from copies of the same unknown state, so channel-simulation lower bounds cannot be transferred to its accuracy dependence without a reduction. The proof supplies that task's quadratic fixed-error lower bound directly.

For the formal previous-output recursion, the angle multiplier is $3^R$ and its optimal initial-copy cost is consequently $\Theta_\eta(9^R)$. If $3^R$ is of order $\sqrt N$, this universal initial-copy benchmark costs order $N$. The construction uses $q$ calls to the known projector reflection; the short formal recursion does not by itself specify a logarithmic-query search implementation. Sources of intermediate states belong to a different resource model and must be charged separately.

The two bounds express different physical scales. The polygon witness first distinguishes an angular harmonic at $M=k$. Reliable discrimination of inputs separated by an angle of order $1/k$, however, requires order $k^2$ copies: their squared $M$-copy overlap is $\cos^{2M}(\pi/(2k))=\exp[-\Theta(M/k^2)]$. A finite collection of copies therefore supplies both a limited angular bandwidth and a limited statistical resolution. A processor can respond nonlinearly to the parameter $\psi$ while remaining linear on its complete input density operator $\hat\rho_\psi^{\otimes M}$. The additional parameter dependence resides in the supplied quantum data, not in access to an ensemble label.

The dependence on accuracy also records which prior information the source supplies. Two complementary results make this role explicit.

\begin{theorem}[Precision under a real-circle promise]\label{thm:circle_precision}
Let $\ket b\in\ker\hat\Pi$ and $\ket g\in\mathrm{ran}\hat\Pi$ be known normalized vectors, including their relative phase convention. For inputs $\ket{\psi_\vartheta}=\cos\vartheta\ket b+\sin\vartheta\ket g$, $\vartheta\in\mathbb R$, let $M_{\rm circ}(k,\eta)$ be the minimum fixed initial-copy budget attaining Eq.~(\ref{eq:implementation_goal}) on this family with unconditional CPTP output. For odd $k\ge3$ and $0<\eta\le1/4$,
\begin{eqnarray}
M_{\rm circ}(k,\eta)=\Theta\!\left(k^2\ln\frac1\eta\right),
\label{eq:circle_precision}
\end{eqnarray}
with constants independent of both $k$ and $\eta$.
\end{theorem}

The complete proof, including uniform constants, is given in~\ref{app:circle}. A block of $k^2$ copies permits an exact heralded conversion with success probability bounded below by a universal constant. Repeating independent blocks and retaining every failure branch gives exponentially decreasing unconditional error. All initial copies are counted; postselection supplies no free resource.

The real-circle promise can arise for a two-level source with a known rotation axis, such as $\ket{\psi_\vartheta}=\hat R_y(2\vartheta)\ket0$, or for linearly polarized photons $\cos\vartheta\ket H+\sin\vartheta\ket V$ in a calibrated polarization basis. The angle is unknown but fixed across the supplied copies, while the basis and relative phase convention are known. The source supplies copies rather than coherent access to the preparation rotation and its inverse. This prior information permits angular components of the copies to be paired coherently in the filters of~\ref{app:circle}.

\begin{proposition}[No universal exact success branch]\label{prop:no_exact_success}
For finite $M$ and odd $k\ge3$, let $\mathcal S$ be an input-independent, trace-nonincreasing completely positive map. If
\begin{eqnarray}
\mathcal S(\hat\rho_\psi^{\otimes M}) = p(\psi)\hat\rho_{A_k(\psi)}
\label{eq:no_exact_success}
\end{eqnarray}
for every pure input in a finite-dimensional space with a nontrivial fixed projector, then $p(\psi)=0$ for every pure input.
\end{proposition}

\begin{proof}---First fix normalized $\ket b\in\ker\hat\Pi$ and $\ket g\in\mathrm{ran}\hat\Pi$ and consider the complex qubit inputs
\begin{eqnarray}
\ket{\psi_z}=\frac{\ket b+z\ket g}{\sqrt{1+|z|^2}}, \quad (z \in \mathbb C).
\end{eqnarray}
Let $\{\hat L_a\}$ be a finite Kraus representation of $\mathcal S$. A positive sum of outer products has rank at most one only if all its nonzero vectors lie on the same ray. Thus, Eq.~(\ref{eq:no_exact_success}) requires every $\hat L_a\ket{\psi_z}^{\otimes M}$ to be collinear with the target or zero.

The components of $\hat L_a(\ket b+z\ket g)^{\otimes M}$ are polynomials in $z$. In a sufficiently small disk about zero, the target has a nonzero $b$ component and component ratio
\begin{eqnarray}
&& \frac{\braket{g}{A_k(\psi_z)}}{\braket{b}{A_k(\psi_z)}} = z h(|z|^2), \nonumber \\
&& h(t)=\frac{\tan(k\arctan\sqrt t)}{\sqrt t} = k + \frac{k(k^2-1)}3t+O(t^2).
\label{eq:nonholomorphic_target}
\end{eqnarray}
The value at $t=0$ is defined by continuity. Shrink the disk so that $h'(t)\ne0$ throughout it. The Wirtinger derivative is
\begin{eqnarray}
\frac{\partial}{\partial\bar z}[z h(|z|^2)] = z^2h'(\abs{z}^2) \ne 0 \quad (z \ne 0).
\end{eqnarray}
If the $b$-component polynomial of one Kraus vector were nonzero, it would be nonzero on an open subset of this punctured disk. There the ratio of its $g$- and $b$-component polynomials would be holomorphic, contradicting the required ratio in Eq.~(\ref{eq:nonholomorphic_target}). The $b$-component polynomial must therefore vanish identically. Collinearity with a target having nonzero $b$ component then forces the entire Kraus vector to vanish on the disk, and the polynomial identity theorem forces it to vanish for every $z$. Continuity includes the state $\ket g$ at $z=\infty$.

This holds for every Kraus operator and every choice of $\ket b,\ket g$. Every pure state with nonzero components in both projector subspaces lies in one such plane; states wholly within either subspace are its endpoints. Therefore, the success probability vanishes for all pure inputs.
\end{proof}

The polynomial structure of a finite quantum program is thus incompatible with the conjugate-amplitude dependence of the universal exact target. Even in a fixed two-level subspace, an unknown relative phase introduces an independent parameter absent from the real-circle promise. Proposition~\ref{prop:no_exact_success} shows that the exact success branches cannot be extended to all such complex inputs. The contrast concerns exact admissibility: it does not prove a separation in approximate accuracy cost. For arbitrary complex pure inputs, the joint bounds remain $\Omega(k^2\ln(1/\eta))$ and $O(k^2/\eta)$ for $0<\eta\le1/4$. Only the fixed-error dependence on $k$ is universally optimal. Prior information enables the sharper construction above without relaxing causal consistency.

\section{Estimation accuracy and preparation resources}\label{sec:learning}

A classical learner produces parameters $\boldsymbol\theta_\psi$ for a circuit $\hat A(\boldsymbol\theta_\psi)$, with estimated pure state $\hat{\rho}_{h_\psi}=\hat A\ketbra{\mathbb0}{\mathbb0}\hat A^\dagger$. Its reflection is $\hat R_{h_\psi}=\hat A\hat R_0\hat A^\dagger$. For pure states,
\begin{eqnarray}
\norm{\hat R_{h_\psi}-\hat R_\psi}_\infty = 2D(\hat{\rho}_{h_\psi},\hat{\rho}_\psi).
\label{eq:reflection_accuracy}
\end{eqnarray}
This connects statistical learning error to coherent control error, but does not imply that a constant learning accuracy preserves the formal amplification law.

\begin{proposition}[Adaptive error and precision allocation]\label{prop:adaptive_control}
Condition on a classical history for which each learner succeeds. Let $\hat\sigma_r$ be the actual pure state, $\hat\rho_r$ the ideal state, and $\hat\tau_r$ the learner's pure estimate. Under the learned-reflection dynamics of Eq.~(\ref{eq:adaptive_updates}), if $D(\hat\tau_r,\hat\sigma_r)\le\epsilon_r$, then $e_r=D(\hat\sigma_r,\hat\rho_r)$ obeys
\begin{eqnarray}
e_{r+1} \le 3e_r+2\epsilon_r,
\quad
e_R \le 3^Re_0+2\sum_{r=0}^{R-1}3^{R-1-r}\epsilon_r.
\label{eq:adaptive_error}
\end{eqnarray}
For $R\ge1$, $e_0=0$, a specified continuous cost model $m_r=a_r/\epsilon_r^2$ with $a_r>0$, and a successful-history error budget $\eta_{\rm good}>0$, the least cost subject to $2\sum_r3^{R-1-r}\epsilon_r\le\eta_{\rm good}$ is
\begin{eqnarray}
\min\sum_r m_r = \frac{4}{\eta_{\rm good}^2}\left(\sum_{r=0}^{R-1}a_r^{1/3}3^{2(R-1-r)/3}\right)^3.
\label{eq:accuracy_allocation}
\end{eqnarray}
\end{proposition}

\begin{proof}---For pure states $\hat\tau,\hat\rho$, the traceless rank-at-most-two operator $\hat\tau-\hat\rho$ has nonzero eigenvalues $\pm\sqrt{1-\tr(\hat\tau\hat\rho)}$. Thus, its operator norm equals its trace distance, proving Eq.~(\ref{eq:reflection_accuracy}). For any state $\hat X$ and unitaries $\hat U,\hat V$, adding and subtracting $\hat V\hat X\hat U^\dagger$ gives $D(\hat U\hat X\hat U^\dagger,\hat V\hat X\hat V^\dagger)\le\norm{\hat U-\hat V}_\infty$. Apply this bound to the updates
\begin{eqnarray}
\hat\sigma_{r+1} = \hat R_{\tau_r}\hat R_\Pi\hat\sigma_r\hat R_\Pi\hat R_{\tau_r},
\quad
\hat\rho_{r+1} = \hat R_{\rho_r}\hat R_\Pi\hat\rho_r\hat R_\Pi\hat R_{\rho_r}.
\label{eq:adaptive_updates}
\end{eqnarray}
First replace $\hat\sigma_r$ by $\hat\rho_r$ at cost $e_r$ by unitary invariance. Replacing the remaining reflection costs at most $2D(\hat\tau_r,\hat\rho_r)\le2(\epsilon_r+e_r)$. This proves the recurrence, and induction gives its sum in Eq.~(\ref{eq:adaptive_error}).

To prove optimality of the allocation, put $w_r=3^{R-1-r}$ and $A=\sum_r a_r^{1/3}w_r^{2/3}$. H\"older's inequality gives
\begin{eqnarray}
A = \sum_r\left(\frac{a_r}{\epsilon_r^2}\right)^{1/3} (w_r\epsilon_r)^{2/3} \le \left(\sum_r m_r\right)^{1/3}\left(\sum_r w_r\epsilon_r\right)^{2/3}.
\label{eq:holder_allocation}
\end{eqnarray}
Cubing and applying the constraint proves the lower bound $\sum_r m_r\ge4A^3/\eta_{\rm good}^2$. It is attained by
\begin{eqnarray}
\epsilon_r=\frac{\eta_{\rm good}}{2A}\left(\frac{a_r}{w_r}\right)^{1/3},
\label{eq:optimal_precision_schedule}
\end{eqnarray}
which saturates both H\"older's inequality and the error constraint. This establishes Eq.~(\ref{eq:accuracy_allocation}).
\end{proof}

If the failure probability at round $r$, conditional on preceding successes, is at most $\delta_r$, then the probability of any failure is at most $\sum_r\delta_r$. Convexity bounds the unconditional error by $\eta_{\rm good}+\sum_r\delta_r$ when $e_0=0$; the failed histories contribute at most their total probability. Rounding each continuous $m_r$ upwards adds less than $R$ copies and preserves the guarantee. These statements assume the routine provides the stated $(\epsilon_r,\delta_r)$ guarantee at that cost, with any confidence dependence included in $a_r$.

Early errors have greater influence on the final state. For constant $a_r=a$, Eq.~(\ref{eq:accuracy_allocation}) grows as $\Theta(a9^R/\eta_{\rm good}^2)$, and the optimal early-round precision is correspondingly finer. This conclusion quantifies how estimation errors affect iterative control. It optimizes the stated estimation-cost model under a sufficient error guarantee; it is not a lower bound on all adaptive quantum strategies. \ref{app:adaptive} treats conditional histories, explicit precision schedules, and the limits of a fixed-accuracy guarantee.

A reusable classical description that reconstructs every pure state to fixed, sufficiently small trace-distance error with constant success probability requires $\Omega(N)$ copies~\cite{haah2016sample,o2016efficient}. Restricted circuit-generated classes can be learned with substantially fewer copies~\cite{zhao2024learning}, but their worst-case lower bounds cannot be transferred to an arbitrary smaller trajectory class. Coherent simulation instead consumes quantum programs and produces the required action without a reusable classical description. In the regime $k^2=o(N)$ at fixed error, Theorem~\ref{thm:optimal_copies} therefore gives a resource separation between universal amplification and universal reconstruction. Both tasks allow all pure inputs; their output requirements differ. This operational separation is the sense in which state-dependent control can require fewer samples than tomography. \ref{app:tomography} derives the reconstruction benchmark and explains how restricted input classes change minimax comparisons. Table~\ref{tab:resource_tasks} summarizes these distinct input promises and output requirements.

\begin{table*}[t]
\begin{adjustbox}{width=1.00\textwidth}
\setlength{\tabcolsep}{10pt}
\begin{tabular}{llll}
\toprule
Task & Input promise & Required output & Copy requirement \\
\midrule
Universal amplification & Arbitrary unknown pure state; known $\hat\Pi$ & One amplified quantum state & $\Theta_\eta(k^2)$ at fixed $0<\eta<1/2$ \\
Real-circle amplification & Known $\ket b,\ket g$ and phase convention; unknown angle & One amplified quantum state & $\Theta(k^2\ln(1/\eta))$ for $0<\eta\le1/4$ \\
Universal reconstruction & Arbitrary unknown pure state & Reusable classical state description & $\Omega(N)$ at fixed small error \\
\bottomrule
\end{tabular}
\end{adjustbox}
\caption{Distinct tasks and their copy requirements. Here $N$ is the state-space dimension, $k\ge3$ is odd, and $\eta$ is the trace-distance error. Amplification has a fixed initial-copy budget and an unconditional output; tomography instead returns a classical estimate with constant success probability. The differing outputs are essential to the comparison.}
\label{tab:resource_tasks}
\end{table*}

The sample comparison above can be made more concrete by including preparation and coherent processing costs. Suppose a source supplies copies only. Let $C_{\rm prep}$ denote the cost of supplying one copy, $C_{\rm ps}$ the cost of one partial-SWAP between the data and program registers, and $C_\Pi$ the cost of the known reflection, all measured in common resource units for a specified implementation model. The construction in Eq.~(\ref{eq:main_constructive_copies}) uses $M=1+q\lceil2\pi^2q/\eta\rceil$ copies, where $q=(k-1)/2$. Its preparation and coherent processing cost is
\begin{eqnarray}
C_{\rm direct}=M C_{\rm prep}+(M-1)C_{\rm ps}+qC_\Pi.
\label{eq:direct_resource_cost}
\end{eqnarray}
Universal reconstruction at fixed small error already requires a worst-case preparation cost $\Omega(NC_{\rm prep})$, before measurement and classical processing are included. At fixed $\eta$, a sufficient asymptotic regime for the direct construction to cost less than this reconstruction step is therefore
\begin{eqnarray}
k^2(C_{\rm prep}+C_{\rm ps})+kC_\Pi=o(NC_{\rm prep}).
\label{eq:resource_advantage_regime}
\end{eqnarray}
For example, $k=3$ and $\eta=0.1$ give $M=199$ and $C_{\rm direct}=199C_{\rm prep}+198C_{\rm ps}+C_\Pi$, independently of $N$ in the copy count. For $N=2^n$, an ideal gate model with arbitrary single-qubit rotations and suitable connectivity implements a register partial-SWAP with $O(n)$ gates using an ancilla and two controlled register swaps. Choosing $\hat\Pi=\ketbra11\otimes\hat{\mathds{1}}_{2^{n-1}}$ makes the known reflection a single-qubit phase flip. This supplies a regime in which coherent processing grows only linearly in $n$, while universal reconstruction retains its dimension-dependent sample cost. The comparison concerns a workflow that first reconstructs the full state. It does not establish optimal gate count or depth, and finite gate synthesis, noise, and repeated reuse of a learned description require their own accounting.

The same accounting principle applies when preparing the input copies requires oracle access. Suppose a complete, successful implementation of an oracle problem has the query upper bound
\begin{eqnarray}
Q_{\rm tot}\le Q_{\rm other}+\sum_r q_r m_r,
\label{eq:query_accounting}
\end{eqnarray}
including all preparation and verification calls, with no uncharged oracle-dependent advice. The standard search lower bound~\cite{bennett1997strengths,zalka1999grover} then requires the right-hand side to be $\Omega(\sqrt N)$. If $q_r=O(1)$, $Q_{\rm other}=O(R)$, and $R=\Theta(\log N)$, this implies an average learning budget $\Omega(\sqrt N/\log N)$. This is a conditional accounting consequence for a successful search implementation, not a universal per-state learning law or a gate-complexity bound. Accounting for the source makes this statement fully consistent with the different initial-copy benchmark above. \ref{app:oracles} details this accounting and the distinction between query and gate costs.

\section{Discussion}\label{sec:discussion}

The copy cost of state-dependent quantum control is governed by both the consistency of the physical operation and the information present in its input. No-signaling requires identical local predictions for equivalent ensemble preparations of that complete input. Within the quantum state space, this requirement yields an inequality that relates the distinguishability of desired outputs to the distinguishability supplied by finitely many copies. For amplitude amplification, an exact ensemble construction identifies a low-copy obstruction, and a two-state comparison establishes the dimension-independent quadratic cost at fixed accuracy. A coherent implementation attains this scaling. Together, the converse and construction turn a causal consistency condition into an operational resource benchmark.

This benchmark clarifies how quantum information enters a computation. A classical estimate can specify a state-dependent circuit because its acquisition consumes physical data. A quantum program can supply the required action directly because the program itself is consumed. An oracle can prepare informative intermediate states because its calls constitute another resource. The costs of these procedures depend on their different inputs and outputs. Our adaptive error bounds quantify how imperfect estimates affect repeated control, while preparation accounting identifies the additional cost when the data source depends on an oracle. These distinctions are necessary for relating a formal control sequence to a complete computational implementation. The stated copy budget includes the active input and every program copy within a fixed maximum budget, with all output branches retained. Preparation costs, gate counts, and circuit depth are separate resources. Supplying a preparation unitary together with its inverse gives coherent access beyond the copy-only model.

Universality in the amplification benchmark means arbitrary pure inputs and a known nontrivial projector. The general finite-scale theorem allows mixed target outputs, but extending amplification to mixed inputs requires a separate definition: $\hat{\mathds{1}}-2\hat\rho$ is generally not a reflection when $\hat\rho^2\ne\hat\rho$. A mixed-input extension must therefore begin by specifying an appropriate target operation and error criterion, which provides a direction for future work. The known real-circle promise yields a joint law in gain and precision; on the full complex pure-state space, the established lower and upper bounds still differ in their dependence on $\eta$. Resolving that gap concerns the conversion of quantum programs into a prescribed output and need not coincide with generic channel simulation. The exact-heralding obstruction makes a further distinction between the real-circle construction and universal complex inputs without asserting a separation in their approximate precision costs. \ref{app:scope} places these results alongside established simulation methods and summarizes the scope of the resource claims.

From a technology perspective, these results provide the resource-design rules for quantum-data workflows in which a quantum feature state or an iteratively learned state serves as a reference for subsequent control. The required copy budget should be set by the output distinctions the task must resolve: in universal pure-state amplification, the initial-copy cost grows quadratically with angular gain at fixed accuracy, independently of the Hilbert-space dimension. When $k^2=o(N)$, directly consuming state copies as a quantum program can require fewer samples for this prescribed single-output task than reconstructing a reusable classical state description. The comparison in Sec.~\ref{sec:learning} identifies conditions under which this advantage persists after preparation and coherent processing costs are included. For learned reflections, the stated stability and estimation-cost model calls for finer precision in earlier rounds; if the reference states depend on an oracle, their preparation and verification calls must also enter the total query cost. These criteria will help assess quantum machine learning and quantum-data architectures in terms of the information physically supplied to the controller and the accuracy its intended action requires~\cite{huang2021information,huang2021power}.

The broader methodological lesson is to connect a computational task to the physical distinctions it must resolve. One first specifies all input resources, then identifies ensembles or nearby states whose desired outputs are well separated, and finally compares the resulting copy bound with an admissible implementation. The replication example shows how the finite-scale theorem applies beyond amplification. The separation from tomography follows from different output requirements: producing one controlled quantum state can require less information than producing a reusable description of its source. Causal consistency and quantum statistical resolution thereby provide complementary parts of the resource analysis of state-dependent control, linking physical admissibility to the information cost of computational action.

\section*{Acknowledgement}

JB and KC was supported by the Korean ARPA-H Project through the Korea Health Industry Development Institute (KHIDI), funded by the Ministry of Health \& Welfare, Republic of Korea (RS-2025-25456722); by the Institute of Information and Communications Technology Planning and Evaluation through a grant funded by the Korean government (RS-2019-II190003, ``Research and Development of Core Technologies for Programming, Running, Implementing and Validating of Fault-Tolerant Quantum Computing System''); and by the Ministry of Trade, Industry and Resources (MOTIR), Korea, under the project ``Industrial Technology Infrastructure Program'' (RS-2024-00466693). We acknowledge the Yonsei University Quantum Computing Project Group for providing support and access to the Quantum System One (Eagle Processor), which is operated at Yonsei University.

\appendix

\section{Operational assumptions and reflection programs}\label{app:setting}\label{sec:SI_setting}

This appendix records the operational assumptions underlying the converse and construction, and distinguishes the prescribed output task from exact programming on independent data.

The input resource is an independently prepared sample of an unknown pure state, $\hat{\rho}_\psi=\ketbra{\psi}{\psi}$, in a Hilbert space of dimension $N\geq2$. The known task specification is a nontrivial orthogonal projector $\hat{\Pi}$, with $0<\mathrm{rank}(\hat{\Pi})<N$. We use the trace distance
\begin{eqnarray}
D(\hat{\rho},\hat{\sigma})=\frac{1}{2}\norm{\hat{\rho}-\hat{\sigma}}_1.
\label{eq:SI_trace_distance}
\end{eqnarray}
It determines the optimal success probability $(1+D)/2$ for distinguishing two states with equal prior probabilities. In particular, orthogonal pure states have distance one. The induced task error is a distance between the actual output density operator and a specified pure output; the output of the physical implementation need not be pure.

For $0<\theta<\pi/2$, write
\begin{eqnarray}
\ket{\psi} &=& \cos\theta\ket{b_\psi}+\sin\theta\ket{g_\psi},
\quad
\sin^2\theta=\bra{\psi}\hat{\Pi}\ket{\psi}, \nonumber \\
\ket{b_\psi} &=& \frac{(\hat{\mathds{1}}-\hat{\Pi})\ket{\psi}}{\cos\theta},
\quad
\ket{g_\psi}=\frac{\hat{\Pi}\ket{\psi}}{\sin\theta}.
\label{eq:SI_components}
\end{eqnarray}
The two components are orthonormal. They are mathematical definitions, not classical information supplied to the processor. For an odd integer $k\geq3$, the desired amplification is
\begin{eqnarray}
\ket{A_k(\psi)} &=& \cos(k\theta)\ket{b_\psi}+\sin(k\theta)\ket{g_\psi}, \nonumber \\
\mathcal{A}_k(\hat{\rho}_\psi) &=& \ketbra{A_k(\psi)}{A_k(\psi)}.
\label{eq:SI_amplification_task}
\end{eqnarray}
At the endpoints, the density operator is defined by continuity: a state entirely inside either subspace is unchanged. Global phases of the input have no effect on the output density operator. We call $k$ the angular gain, since the target population need not increase after an overshoot.

The converse and implementation results use different assumptions. For the converse, a deterministic device is defined on every state of its $M$ input registers and returns a quantum output state. Ordinary probabilistic averaging and remote ensemble preparation have their usual quantum meaning. Averaging here concerns the outputs conditional on an actual preparation procedure; no-signaling requires agreement between different remotely selectable decompositions of the same input density operator. No affine or CPTP extension is assumed in advance. Accuracy is promised only on the specified pure tensor powers. The auxiliary states used below need not satisfy that promise; the full input domain and normalized outputs are necessary for the causal proof.

For quantum achievability, a deterministic implementation is a completely positive trace-preserving (CPTP) map $\mathcal{L}_{M,k}$ such that
\begin{eqnarray}
\sup_{\ket{\psi}}D\!\left(\mathcal{L}_{M,k}(\hat{\rho}_\psi^{\otimes M}),\mathcal{A}_k(\hat{\rho}_\psi)\right)\leq\eta.
\label{eq:SI_task_error}
\end{eqnarray}
The integer $M$ counts \emph{all} initial-state copies, including an active output register. Ancillas independent of $\psi$, collective operations, measurements, feedforward, and internal randomness are allowed. The total available-copy budget is fixed, and the final output is unconditional. Heralded intermediate steps are allowed only when their success and failure branches are both included in the final channel and resource count.

Completing unequal input density operators to decompositions of a common state gives the quantitative no-signaling inequality below. CPTP implementations satisfy the same inequality by trace-distance contraction. We do not derive complete positivity or the quantum state space from no-signaling; we identify the causal constraint within that state space and then exhibit quantum channels that attain the asserted scaling.

This is a task-oriented use of quantum samples. It does not require a classical estimate of every amplitude or a reusable preparation circuit. Consequently, the dimension-independent upper bound in Theorem~\ref{thm:optimal_copies} does not provide dimension-independent full-state tomography. Likewise, a coherent preparation unitary supplied together with its inverse constitutes a different resource from copies of its output. Keeping these distinctions explicit is necessary when interpreting the sample count as a physical computational cost.

\subsection{Reflection geometry and exact programmability}\label{sec:SI_geometry}

Define the reflections
\begin{eqnarray}
\hat{R}_\psi=\hat{\mathds{1}}-2\hat{\rho}_\psi,
\quad
\hat{R}_\Pi=\hat{\mathds{1}}-2\hat{\Pi},
\quad
\hat{G}_\psi=-\hat{R}_\psi\hat{R}_\Pi.
\label{eq:SI_reflections}
\end{eqnarray}
For a \emph{fixed} state $\psi$, both reflections leave the span of $\ket{b_\psi},\ket{g_\psi}$ invariant. In that ordered basis,
\begin{eqnarray}
\hat{R}_\psi = \left( \begin{array}{cc} -\cos2\theta & -\sin2\theta \\ -\sin2\theta & \cos2\theta \end{array} \right),
\quad
\hat{G}_\psi = \left( \begin{array}{cc} \cos2\theta & -\sin2\theta \\ \sin2\theta & \cos2\theta \end{array} \right).
\label{eq:SI_rotation_matrix}
\end{eqnarray}
It follows directly that
\begin{eqnarray}
\hat{G}_\psi^q\ket{\psi} = \cos[(2q+1)\theta]\ket{b_\psi} + \sin[(2q+1)\theta]\ket{g_\psi}.
\label{eq:SI_fixed_reflection}
\end{eqnarray}
Thus, $q=(k-1)/2$ applications implement the mathematical target in Eq.~(\ref{eq:SI_amplification_task}) if the initial-state reflection is available. The restriction to odd $k$ is natural for this two-reflection construction, as used in Sec.~\ref{sec:model}.

The same calculation explains the previous-output construction. In a fixed plane, let $\ket{\psi_\vartheta}=\cos\vartheta\ket{b}+\sin\vartheta\ket{g}$, where $\vartheta$ may be any real number. Then
\begin{eqnarray}
-\hat{R}_{\psi_\vartheta}\hat{R}_\Pi\ket{\psi_\vartheta}=\ket{\psi_{3\vartheta}}.
\label{eq:SI_tripling}
\end{eqnarray}
The signed angle avoids an artificial restriction to the first quadrant. Repeating the state-dependent update $R$ times gives the formal target $\hat{\rho}_{\psi_{3^R\vartheta}}$. For $\vartheta=\arcsin(1/\sqrt{N})$, a constant target population is reached after $O(\log N)$ such formal updates. For example, stopping at the first angle above a fixed $c<\pi/6$ places that angle in $[c,3c)$, so its success probability is at least $\sin^2c$. This is a statement about a nonlinear prescription indexed by the unknown input, not a query algorithm with a specified physical implementation. The missing operation is the reflection about the unknown current state.

The following standard programmability argument supplies a precise obstruction~\cite{PhysRevLett.79.321,kumar2011quantum}.

\begin{proposition}[No exact finite-copy reflection program]\label{prop:SI_no_reflection}
No deterministic processor can use $m<\infty$ copies of an arbitrary unknown state $\ket{\psi}$ to implement the unitary channel of $\hat{R}_\psi$ exactly on every independent data input.
\end{proposition}
\begin{proof}
Suppose an isometric dilation $\hat{V}$ of such a processor existed. Exactness of its unitary output channel implies
\begin{eqnarray}
\hat{V}\bigl(\ket{\psi}^{\otimes m}\ket{x}\bigr)=\ket{e_\psi}\otimes\hat{R}_\psi\ket{x}
\label{eq:SI_program_isometry}
\end{eqnarray}
for every data vector $\ket{x}$. The environment vector is independent of $x$: each pure output factorizes from the environment, and applying linearity to superpositions of two data basis states forces their environment factors to agree, with any common phase absorbed into $\ket{e_\psi}$. The environment may contain the consumed program and measurement records; the proof does not require preserving the program.

Preservation of inner products for two program states and arbitrary data vectors gives
\begin{eqnarray}
\braket{\psi}{\phi}^{m}\braket{x}{y}=\braket{e_\psi}{e_\phi}\bra{x}\hat{R}_\psi^\dagger\hat{R}_\phi\ket{y}.
\label{eq:SI_program_overlap}
\end{eqnarray}
Choose nonorthogonal states whose reflections are not equal up to phase, for instance $\ket{\psi}=\ket{0}$ and $\ket{\phi}=(\ket{0}+\ket{1})/\sqrt{2}$. The left-hand coefficient is nonzero for every finite $m$, so the environment overlap cannot vanish. Eq.~(\ref{eq:SI_program_overlap}) would then require $\hat{R}_\psi^\dagger\hat{R}_\phi$ to be a scalar multiple of the identity, a contradiction.
\end{proof}

The proposition concerns exact operation on arbitrary independent data. It does not, by itself, give a quantitative lower bound for the correlated-input task of Eq.~(\ref{eq:SI_task_error}). That task is analyzed directly in Secs.~\ref{sec:causality} and \ref{sec:optimal}. Neither result excludes an exact reflection compiled from an already known classical state description; the cost of obtaining that description is a separate question.

\section{Remote preparation and finite-copy ensembles}\label{app:ensembles}\label{sec:SI_ensembles}

Choose any normalized $\ket{b}\in\ker\hat{\Pi}$ and $\ket{g}\in\mathrm{ran}\hat{\Pi}$. Their span is a valid two-dimensional restriction in every allowed dimension. With $\hat{\rho}_\vartheta=\ketbra{\psi_\vartheta}{\psi_\vartheta}$, define the two polygon ensembles
\begin{eqnarray}
\mathcal{E}_\alpha = \left\{\frac{1}{k},\hat{\rho}_{\alpha+j\pi/k}:j=0,\ldots,k-1\right\}, \quad \alpha_0 = 0, \quad \alpha_1=\frac{\pi}{2k}.
\label{eq:SI_polygon_ensembles}
\end{eqnarray}
Eq.~(\ref{eq:SI_fixed_reflection}), applied to signed angles, shows that every member of $\mathcal{E}_{\alpha_0}$ has target density operator $\ketbra{b}{b}$, whereas every member of $\mathcal{E}_{\alpha_1}$ has target $\ketbra{g}{g}$. The factors $(-1)^j$ in their state vectors are unobservable global phases.

\subsection{Explicit remote preparation for one copy} 

For either ensemble, the mean input is $\hat{\mathds{1}}_2/2$. Share the entangled state $\ket{\Phi}= (\ket{b}_A\ket{b}_B+\ket{g}_A\ket{g}_B)/\sqrt{2}$. Bob's positive operator-valued measure (POVM) has effects
\begin{eqnarray}
\hat{E}^{(\alpha)}_j=\frac{2}{k}\hat{\rho}_{\alpha+j\pi/k}^{T},
\quad
\sum_{j=0}^{k-1}\hat{E}^{(\alpha)}_j=\hat{\mathds{1}}_2.
\label{eq:SI_steering_povm}
\end{eqnarray}
The transpose is in the displayed Schmidt basis. Since the vectors here are real, it leaves the matrices unchanged. Bob's outcome $j$ occurs with probability $1/k$ and prepares Alice's conditional state $\hat{\rho}_{\alpha+j\pi/k}$, because
\begin{eqnarray}
\tr_B\!\left[(\hat{\mathds{1}}\otimes\hat{E}^{(\alpha)}_j)\ketbra{\Phi}{\Phi}\right] = \frac{1}{2}(\hat{E}^{(\alpha)}_j)^T = \frac{1}{k}\hat{\rho}_{\alpha+j\pi/k}.
\end{eqnarray}
Bob does not transmit his outcome. A device realizing the ideal target would nevertheless convert his choice of POVM into the distinguishable states $\ketbra{b}{b}$ and $\ketbra{g}{g}$ at Alice's location.

For the tripling map $k=3$, these are two rotated trine POVMs. Their effects are explicitly
\begin{eqnarray}
\{\hat{E}^{(0)}_j\} &=&
\left\{
\frac{2}{3} \left(\begin{array}{cc} 1 & 0 \\ 0 & 0 \end{array}\right),
\frac{1}{6} \left(\begin{array}{cc} 1 & \sqrt{3} \\ \sqrt{3} & 3 \end{array}\right),
\frac{1}{6} \left(\begin{array}{cc} 1 & -\sqrt{3} \\ -\sqrt{3} & 3 \end{array}\right)
\right\}, \nonumber \\
\{\hat{E}^{(\pi/6)}_j\} &=&
\left\{
\frac{1}{6} \left(\begin{array}{cc} 3 & \sqrt{3} \\ \sqrt{3} & 1 \end{array}\right),
\frac{2}{3} \left(\begin{array}{cc} 0 & 0 \\ 0 & 1 \end{array}\right),
\frac{1}{6} \left(\begin{array}{cc} 3 & -\sqrt{3} \\ -\sqrt{3} & 1 \end{array}\right)
\right\}.
\label{eq:SI_trine_povms}
\end{eqnarray}
Each set sums to the identity and consists of positive rank-one effects. This supplies a direct signaling test for the nonlinear target without identifying an entangled-state transpose identity with a remotely applied oracle.

\subsection{Equality of the complete input states for \texorpdfstring{$M<k$}{M < k}} 

In a sample-assisted experiment, one must compare the complete input density operators
\begin{eqnarray}
\hat{\Omega}^{(M)}_\alpha=\frac{1}{k}\sum_{j=0}^{k-1}\hat{\rho}_{\alpha+j\pi/k}^{\otimes M}.
\label{eq:SI_Mcopy_ensemble}
\end{eqnarray}
This is a mixture of independently prepared copies \emph{conditional on the chosen state}; it is not the tensor power of the one-copy mixture. Confusing these expressions would give an incorrect signaling conclusion for unrestricted $M$.

Theorem~\ref{thm:causal_threshold} proves that these two complete input states coincide for $M<k$, by resolving their angular Fourier orders. Their common output must then lie within distance $\eta$ of each of the two orthogonal target states, implying $\eta\ge1/2$.

The equality of the complete input states makes this an ensemble-consistency obstruction. Its remote preparation follows explicitly from the Hughston--Jozsa--Wootters construction~\cite{hughston1993complete}. Purify the common state as $\ket{\Omega}=\sum_\ell\sqrt{\lambda_\ell}\ket{\ell}_A\ket{\ell}_B$, on its support, and take
\begin{eqnarray}
\hat{F}^{(\alpha)}_j = \left[\hat{\Omega}^{-1/2}\frac{\hat{\rho}_{\alpha+j\pi/k}^{\otimes M}}{k}\hat{\Omega}^{-1/2}\right]^T.
\label{eq:SI_Mcopy_steering}
\end{eqnarray}
The effects sum to the support identity, and their conditional states at Alice are exactly the desired product states with probability $1/k$. The inverse is only taken on the support. Thus, no assumption of remote cloning is involved. In dimension two, the channel that discards its input and outputs $\hat{\mathds{1}}_2/2$ attains error $1/2$ for every pure target, making the threshold exact when $M<k$. This tightness statement is restricted to dimension two.

\subsection{From no-signaling to a quantitative ensemble inequality} 

The equal-input test extends to unequal inputs by a direct completion of ensembles. Consider two input ensembles with mean density operators $\hat{\Omega}_0,\hat{\Omega}_1$, actual mean outputs $\hat{\Phi}_0,\hat{\Phi}_1$, and desired mean outputs $\hat{\Gamma}_0,\hat{\Gamma}_1$. Let $\eta_a$ bound the ensemble-averaged output error for ensemble $a$, so convexity gives $D(\hat{\Phi}_a,\hat{\Gamma}_a)\leq\eta_a$.

Under the quantum state-space and device-domain assumptions of~\ref{app:setting}, the completion argument in Lemma~\ref{lem:ensemble_consistency} applies separately to the two average errors and gives
\begin{eqnarray}
D(\hat{\Gamma}_0,\hat{\Gamma}_1) \leq D(\hat{\Omega}_0,\hat{\Omega}_1)+\eta_0+\eta_1.
\label{eq:SI_causal_ensemble}
\end{eqnarray}
Indeed, the same Jordan completion bounds the distance between the actual mean outputs by $D(\hat\Omega_0,\hat\Omega_1)$, and the triangle inequality adds the two errors $\eta_0$ and $\eta_1$.

The auxiliary pure states may be entangled across the $M$ input registers. This is why the device must have an action outside its promised pure-tensor-power inputs. A standard quantum processor has such an action automatically, and its CPTP contraction gives a shorter equivalent necessary inequality. The completion argument identifies the no-signaling content explicitly while making no claim that the principle, without the quantum state geometry and input-domain assumptions, establishes this bound in a general probabilistic theory.

\subsection{Exact distinguishability of the rotated polygon inputs} 

For completeness, the same ensembles give a computable bound for every $M$. Define
\begin{eqnarray}
b_\ell = 2^{-M}{M \choose \ell},
\quad
p_a = {\!\!\!\!\!\!\!\!} \sum_{0 \leq \ell \leq M \atop \ell \equiv a \, ({\rm mod} \, 2k)} {\!\!\!\!} b_\ell, \quad (0 \leq a < 2k).
\end{eqnarray}
Then,
\begin{eqnarray}
B_{M,k} := D(\hat{\Omega}^{(M)}_{\alpha_0},\hat{\Omega}^{(M)}_{\alpha_1}) = 2\sum_{r=0}^{k-1}\sqrt{p_rp_{r+k}},
\quad
\eta \geq \frac{1-B_{M,k}}{2}.
\label{eq:SI_polygon_bias}
\end{eqnarray}
Proposition~\ref{prop:finite_bias} proves this equality by decomposing the symmetric input space into residue blocks modulo $k$.

For the inequality in Eq.~(\ref{eq:SI_polygon_bias}), Eq.~(\ref{eq:SI_causal_ensemble}) gives $1\leq2\eta+B_{M,k}$ because the two desired mean outputs are orthogonal. For CPTP implementations, the same statement follows directly from contraction on the two mean inputs. At $M=k=3$, $B_{3,3}=1/4$, so every tripling implementation has $\eta\geq3/8$. These are rigorous lower bounds, not a claim that the polygon construction solves the general minimax problem.

There is an exact special case at the threshold, provided the input is promised to lie on the \emph{known real great circle} $\ket{\psi_\vartheta}=\cos\vartheta\ket{b}+\sin\vartheta\ket{g}$. At $M=k$, set $\ket{y_\pm}=(\ket b\pm i\ket g)/\sqrt2$ and define the operator from the input registers to the output qubit by
\begin{eqnarray}
\hat{K}=\ket{y_+}\bra{y_+}^{\otimes k}+\ket{y_-}\bra{y_-}^{\otimes k}.
\end{eqnarray}
Its two input vectors are orthogonal, so $\hat{K}^\dagger\hat{K}\leq\hat{\mathds{1}}$. Eq.~(\ref{eq:symmetric_expansion}) gives
\begin{eqnarray}
\hat{K}\ket{\psi_\vartheta}^{\otimes k}=2^{(1-k)/2}\ket{\psi_{k\vartheta}}.
\end{eqnarray}
Include the complementary outcome by defining the deterministic channel
\begin{eqnarray}
\mathcal{C}(\hat{X})=\hat{K}\hat{X}\hat{K}^\dagger + \tr[(\hat{\mathds{1}}-\hat{K}^\dagger\hat{K})\hat{X}]\frac{\hat{\mathds{1}}_2}{2}.
\label{eq:SI_circle_channel}
\end{eqnarray}
Its output is $2^{1-k}\hat{\rho}_{k\vartheta}+(1-2^{1-k})\hat{\mathds{1}}_2/2$, with trace error $1/2-2^{-k}$ for every promised input. Since $B_{k,k}=2^{1-k}$, the polygon lower bound is saturated. The restricted-circle minimax error is therefore $1/2$ for $M<k$ and $1/2-2^{-k}$ for $M=k$. Both branches of Eq.~(\ref{eq:SI_circle_channel}) are retained, and all $k$ copies are counted. This does not assert the same optimum for an arbitrary complex qubit state.

\section{Finite-scale separation and state replication}\label{app:finite_scale}\label{sec:SI_lower}

Theorem~\ref{thm:finite_scale} converts a finite output separation into a copy requirement. The following counterexample explains why a large derivative alone is insufficient, and the replication application illustrates the same principle for a different output task.

\subsection{Why a differential expansion alone is insufficient} 

At fixed additive output error, a large local derivative need not produce an observable separation. The following construction makes this distinction explicit over the entire qubit pure-state space. Fix $0<\epsilon<\min\{\eta,\pi/4\}$, let $n$ be a positive integer, and set $\hat Z=\ketbra{0}{0}-\ketbra{1}{1}$. Define
\begin{eqnarray}
f_n(\hat\rho) &=& \epsilon\sin[n\,\tr(\hat\rho\hat Z)], \nonumber\\
\ket{t_n(\rho)} &=& \cos f_n(\hat\rho)\ket0+\sin f_n(\hat\rho)\ket1, \nonumber\\
T_n(\rho) &=& \ketbra{t_n(\rho)}{t_n(\rho)}.
\label{eq:SI_small_range_target}
\end{eqnarray}
This prescription depends only on the density operator and is independent of its global phase. Along the input curve
\begin{eqnarray}
\ket{\psi_\theta}=\cos(\pi/4+\theta)\ket0+\sin(\pi/4+\theta)\ket1,
\end{eqnarray}
the input Fubini--Study speed is one, $\tr(\hat\rho_{\psi_\theta}\hat Z)=-\sin2\theta$, and $df_n(\hat\rho_{\psi_\theta})/d\theta|_{\theta=0}=-2\epsilon n$. The local output Fubini--Study speed is consequently $2\epsilon n$, which is unbounded as $n$ grows. Nevertheless,
\begin{eqnarray}
\sup_\psi D\bigl(T_n(\hat\rho_\psi),\ketbra0{0}\bigr) = \sup_\psi\abs{\sin f_n(\hat\rho_\psi)} \le \sin\epsilon<\eta.
\label{eq:SI_small_range_approximation}
\end{eqnarray}
The constant-output channel therefore meets the error requirement without using input information. A finite-scale separation exceeding $2\eta$, as required in Theorem~\ref{thm:finite_scale}, rules out precisely this possibility. Differential growth is informative only when additional control of its range or regularity guarantees the required finite output separation.

\subsection{Application to unknown pure-state replication} 

Replication supplies a task distinct from amplitude amplification. The target is the entire $r$-register state $T_r(\psi)=\hat\rho_\psi^{\otimes r}$, and error is measured globally on these registers, rather than on their individual reduced states.

\begin{corollary}[Linear copy cost for global replication]\label{cor:SI_replication}
For every dimension $N\ge2$ and fixed $0\le\eta<1/2$, an unconditional implementation of the replication target on all pure inputs requires $M=\Omega_\eta(r)$ under the assumptions of Theorem~\ref{thm:finite_scale}. Keeping $r$ input copies implements the task exactly, so its minimum quantum copy cost is $\Theta_\eta(r)$. At $\eta=0$, the minimum is exactly $r$.
\end{corollary}
\begin{proof}
Choose any $c>0$ such that $\sqrt{1-e^{-c}}>2\eta$. Such a finite $c$ exists whenever $\eta<1/2$. In any two-dimensional input subspace choose
\begin{eqnarray}
\ket{\psi} &=& \ket0, \nonumber \\
\ket{\phi_r} &=& e^{-c/(2r)}\ket0+\sqrt{1-e^{-c/r}}\ket1.
\label{eq:SI_replication_pair}
\end{eqnarray}
Their input squared overlap is $e^{-c/r}$, and their target separation is the constant $\Delta_c=\sqrt{1-e^{-c}}$. The pure-state distance formula and the causal pair bound give
\begin{eqnarray}
\Delta_c-2\eta \le \sqrt{1-e^{-cM/r}},
\quad
M \ge \frac{r}{c}\ln\frac{1}{1-(\sqrt{1-e^{-c}}-2\eta)^2}.
\label{eq:SI_replication_lower}
\end{eqnarray}
The coefficient of $r$ is positive and depends only on the chosen fixed error and $c$, proving the lower bound. Their Fubini--Study separation satisfies $\delta_r=\arccos(e^{-c/(2r)})=\sqrt{c/r}[1+O(r^{-1})]$, making explicit the scale $L=\sqrt r$ in Theorem~\ref{thm:finite_scale}. The upper bound follows from the identity channel on $r$ supplied copies. For $\eta=0$, Eq.~(\ref{eq:SI_replication_lower}) reduces exactly to $M\ge r$, because its logarithm is $c$. This also proves the exact claim.
\end{proof}

This example illustrates the scope of the finite-scale bound; it is not claimed as a new cloning theorem. In both replication and amplification, a transformation exposes distinctions hidden at the single-copy input scale. No-signaling requires those distinctions to be present in the complete physical input, and the quantum overlap law quantifies the number of samples needed to supply them.

\section{Joint gain--accuracy scaling on a known real great circle}\label{app:circle}\label{sec:SI_circle_accuracy} 

The accuracy dependence can be resolved under the explicit promise that $\ket b$ and $\ket g$ are known and the input is $\ket{\psi_\vartheta}=\cos\vartheta\ket b+\sin\vartheta\ket g$, with arbitrary real $\vartheta$. Define $M_{\rm circ}(k,\eta)$ as the least fixed initial-copy budget for this promised task, retaining the deterministic, unconditional output requirement. The promise supplies a known plane and phase convention, rather than a classical description of the unknown angle.

We prove Theorem~\ref{thm:circle_precision} by establishing the following explicit bounds for odd $k\ge3$ and $0<\eta\le1/4$. With $s_0=e^{-9/4}/2$,
\begin{eqnarray}
\frac{3k^2}{5\pi^2}\ln\frac1\eta \le M_{\rm circ}(k,\eta) \le k^2\left\lceil\frac{\ln[1/(2\eta)]}{-\ln(1-s_0)}\right\rceil.
\label{eq:SI_circle_quantitative}
\end{eqnarray}

\begin{proof}[Proof of Theorem~\ref{thm:circle_precision}]
The two-state lower bound applies to the promised inputs at angles $0$ and $\pi/(2k)$. Hence,
\begin{eqnarray}
M_{\rm circ}(k,\eta)\ge
\frac{\ln[1/(4\eta(1-\eta))]}{-\ln\cos^2[\pi/(2k)]}.
\end{eqnarray}
For $x=\pi/(2k)\le\pi/6$,
\begin{eqnarray}
-\ln\cos^2x = 2\int_0^x\tan t\,dt \le x^2\sec^2x\le\frac{\pi^2}{3k^2}.
\end{eqnarray}
Here $\tan t=\int_0^t\sec^2u\,du\le t\sec^2x$. Also $4\eta^{4/5}(1-\eta)$ increases on $(0,1/4]$ and is less than one at $1/4$, giving $\ln[1/(4\eta(1-\eta))]\ge\frac15\ln(1/\eta)$. These inequalities prove the lower bound in Eq.~(\ref{eq:SI_circle_quantitative}).

For the upper bound, use blocks of $n=k^2=2m+1$ copies. Set $\ket{y_\pm}=(\ket b\pm i\ket g)/\sqrt2$, and let $\ket\ell$ be the normalized symmetric vector containing $\ell$ factors $\ket{y_-}$. With $b_\ell=2^{-n}{n \choose \ell}$,
\begin{eqnarray}
\ket{\psi_\vartheta}^{\otimes n} = e^{-in\vartheta}\sum_{\ell=0}^n\sqrt{b_\ell}\,e^{2i\ell\vartheta}\ket\ell.
\label{eq:SI_circle_block_input}
\end{eqnarray}
Pair the $k$ indices $m-k+1\le\ell\le m$ with $\ell+k$, and define $c_\ell=\min(b_\ell,b_{\ell+k})$. The success Kraus operators are
\begin{eqnarray}
\hat K_\ell = \sqrt{\frac{c_\ell}{b_\ell}}\ket{y_+}\bra\ell + \sqrt{\frac{c_\ell}{b_{\ell+k}}}\ket{y_-}\bra{\ell+k}.
\label{eq:SI_circle_success_kraus}
\end{eqnarray}
All index pairs are disjoint, so $\sum_\ell\hat K_\ell^\dagger\hat K_\ell\le\hat{\mathds{1}}$. The complementary positive effect defines a failure outcome; this completes the instrument on the full input space. Direct substitution gives
\begin{eqnarray}
\hat K_\ell\ket{\psi_\vartheta}^{\otimes n} = \sqrt{2c_\ell}\,e^{i(2\ell-n+k)\vartheta}\ket{\psi_{k\vartheta}}.
\label{eq:SI_circle_exact_success}
\end{eqnarray}
Each success outcome therefore prepares the exact target, and the total success probability $s=2\sum_\ell c_\ell$ is independent of $\vartheta$.

We now bound $s$ below uniformly in $k$. A binomial variable $L$ with probabilities $b_\ell$ has mean $n/2$ and variance $k^2/4$. Chebyshev's inequality gives $\Pr(|L-n/2|<k)\ge3/4$. This interval contains $2k$ integers, and $b_m$ is maximal, so $b_m\ge3/(8k)\ge1/(4k)$. By symmetry and unimodality, every selected coefficient is at least $b_{m+k}$. Furthermore,
\begin{eqnarray}
\fl \frac{b_{m+k}}{b_m} = \prod_{r=1}^k\frac{m+2-r}{m+r} \ge \exp\!\left[-\sum_{r=1}^k\frac{2r-2}{m+2-r}\right] \ge \exp\!\left[-\frac{k(k-1)}{m+2-k}\right] \ge e^{-9/4}.
\label{eq:SI_circle_binomial_bound}
\end{eqnarray}
The first inequality uses $\ln(1-x)\ge-x/(1-x)$ for $0\le x<1$. The last is equivalent to $2k(k-1)/(k^2-2k+3)\le9/4$, which follows from $(k-5)^2+2\ge0$. Thus, $s\ge2k e^{-9/4}/(4k)=s_0$.

Apply the instrument to $r$ independent blocks and retain the first successful output. If all blocks fail, output $\hat{\mathds{1}}_2/2$. The unconditional state is
\begin{eqnarray}
\hat\sigma_\vartheta^{(r)} = [1-(1-s)^r]\hat\rho_{\psi_{k\vartheta}} + (1-s)^r\frac{\hat{\mathds{1}}_2}{2}.
\label{eq:SI_circle_unconditional}
\end{eqnarray}
Its error is exactly $(1-s)^r/2$. Choosing $r=\lceil\ln[1/(2\eta)]/[-\ln(1-s_0)]\rceil$ proves the upper bound. All $rk^2$ initial registers are included in the fixed budget; unused registers after an earlier success may be discarded. The final task is deterministic and retains every failure branch.
\end{proof}

The same angular Fourier representation that gives the causal obstruction therefore also supplies an achievable construction once enough frequencies have appreciable weight. Internal heralding does not evade no-signaling: the unconditional output includes the all-failure event. The theorem resolves accuracy dependence only for the stated real-circle promise. For $0<\eta\le1/4$, the universal complex-input problem retains the bounds $\Omega(k^2\ln(1/\eta))$ and $O(k^2/\eta)$.

\section{Conditional learning and adaptive accuracy}\label{app:adaptive}\label{sec:SI_learning}

The coherent construction consumes copies as a quantum program. A controller that instead estimates a classical circuit description requires the conditional accuracy guarantees described here. The learner receives independent copies of a pure state and returns a classical circuit description. Given a learned preparation circuit $\hat A(\boldsymbol\theta)$, let
\begin{eqnarray}
\ket{h}=\hat A(\boldsymbol\theta)\ket{\mathbb0},
\quad
\hat R_h=\hat A(\boldsymbol\theta)\hat R_0\hat A(\boldsymbol\theta)^\dagger,
\quad
\hat R_0=\hat{\mathds{1}}-2\ketbra{\mathbb0}{\mathbb0}.
\label{eq:SI_learned_reflection}
\end{eqnarray}
Here $\ket{\mathbb0}=\ket0^{\otimes n}$ for $N=2^n$. Once the classical description is supplied, this is a conventional unitary circuit. Its acquisition cost and its subsequent gate count are distinct resources.

\subsection{An exact relation between state and reflection errors} 

The identity in Eq.~(\ref{eq:reflection_accuracy}), proved in Proposition~\ref{prop:adaptive_control}, converts trace-distance estimation error into operator-norm reflection error. It applies to all pure states, including identical and orthogonal inputs.

For any normalized vector $\ket v$, the trace distance between the two pure outputs $\hat U\ket v$ and $\hat V\ket v$ is at most $\norm{\hat U-\hat V}_\infty$. Thus, a trace-distance learning error $\epsilon$ gives an output error at most $2\epsilon$ for a single reflection. The statement is about the specified Householder representatives; arbitrary unitary channels are physically insensitive to an overall phase.

\subsection{Conditional pure trajectories} 

An adaptive learning routine produces random classical transcripts. Conditional on the full preceding transcript, the actual state prepared by a unitary circuit is pure. Denote its density operator by $\hat\sigma_r$, the estimate returned by the next learner by $\hat\tau_r$, and the ideal nonlinear trajectory by $\hat\rho_r$. Set
\begin{eqnarray}
e_r=D(\hat\sigma_r,\hat\rho_r),
\quad
D(\hat\tau_r,\hat\sigma_r)\le\epsilon_r.
\label{eq:SI_conditional_error}
\end{eqnarray}
The ideal and actual updates are, respectively,
\begin{eqnarray}
\hat\rho_{r+1}=\hat R_{\rho_r}\hat R_\Pi\hat\rho_r\hat R_\Pi\hat R_{\rho_r},
\quad
\hat\sigma_{r+1}=\hat R_{\tau_r}\hat R_\Pi\hat\sigma_r\hat R_\Pi\hat R_{\tau_r}.
\label{eq:SI_two_updates}
\end{eqnarray}
The global phase in the vector update does not appear in these density operators. The copies supplied to a round must be independent copies of the conditional actual state. Averaging over a shared classical history can correlate them, so an unconditional mixture must not silently be replaced by a tensor power of its one-copy marginal.

Proposition~\ref{prop:adaptive_control} proves the recurrence and its iterated form in Eq.~(\ref{eq:adaptive_error}) on every history satisfying Eq.~(\ref{eq:SI_conditional_error}) at each round.

The factor three is the local sensitivity of angle tripling: nearby real-plane inputs with separation $d\theta$ have initial trace distance $|d\theta|+o(d\theta)$ and output distance $3|d\theta|+o(d\theta)$. It is therefore not an artefact of an unnecessarily large continuity constant. More generally, the ideal odd-$k$ target is $k$-Lipschitz on pure states. Indeed, compare $q$ fixed-reflection iterates on two inputs, insert one differing reflection at a time, and use Eq.~(\ref{eq:reflection_accuracy}). The input contribution is $D(\rho,\sigma)$, and the $q$ reflection contributions sum to at most $2qD(\rho,\sigma)$, giving $(2q+1)D(\rho,\sigma)$.

If the learner at round $r$ has conditional failure probability at most $\delta_r$ for every preceding successful history, the probability that any round fails is at most $\sum_r\delta_r$. On the complement, Eq.~(\ref{eq:adaptive_error}) holds. If the right-hand side is at most $\eta_{\rm good}$, the unconditioned output error is at most $\eta_{\rm good}+\sum_r\delta_r$, by convexity and the unit upper bound on trace distance. These two errors can be budgeted separately. Additional circuit errors can be added to the one-step bound as output-distance terms; they must not be absorbed into a statistical accuracy without specification.

\subsection{A precision schedule with a proven cost guarantee} 

Assume $R\ge1$, $e_0=0$, $\eta_{\rm good}>0$, and a stated continuous learning-cost model
\begin{eqnarray}
m_r=\frac{a_r}{\epsilon_r^2}, \quad (a_r>0),
\label{eq:SI_cost_model}
\end{eqnarray}
where any fixed confidence factors have already been included in $a_r$. Such a model can describe a specified routine, or a simplified scaling regime of a tomography guarantee. It is not an assumption that every restricted state family has this cost. Define
\begin{eqnarray}
w_r=3^{R-1-r},
\quad
A=\sum_{r=0}^{R-1}a_r^{1/3}w_r^{2/3}.
\label{eq:SI_schedule_weights}
\end{eqnarray}

For positive precisions obeying $2\sum_r w_r\epsilon_r\le\eta_{\rm good}$, Proposition~\ref{prop:adaptive_control} proves the minimum in Eq.~(\ref{eq:accuracy_allocation}) and the attaining schedule in Eq.~(\ref{eq:optimal_precision_schedule}). This is the optimization of the specified continuous cost under the stated stability certificate.

For a uniform coefficient $a_r=a$, write
\begin{eqnarray}
S_R=\sum_{r=0}^{R-1}3^{2(R-1-r)/3}=\frac{3^{2R/3}-1}{3^{2/3}-1}.
\end{eqnarray}
Then $\epsilon_r=\eta_{\rm good}w_r^{-1/3}/(2S_R)$ and the certified cost is $4aS_R^3/\eta_{\rm good}^2=\Theta(a9^R/\eta_{\rm good}^2)$. Integer budgets $m_r=\lceil a_r/\epsilon_r^2\rceil$ add fewer than $R$ copies. For $0<\eta_{\rm good}<1/2$ the displayed schedule has $\epsilon_r<1/4$, so it lies in the usual small-error regime. If a routine has additional accuracy-dependent logarithms, retain them in the cost function or upper-bound them over the chosen schedule; an exact optimum for the simplified model must not be asserted for a different cost law.

This is an achievable resource allocation once the conditional intermediate sources and the corresponding learning guarantees exist. It optimizes a sufficient stability certificate. It does not prove that every successful adaptive protocol must follow the same schedule, nor that the initial-copy cost equals the sum of intermediate-source copies.

\subsection{Why fixed accuracy alone does not certify fast amplification} 

For the usual search initialization, $\theta_0=\arcsin(1/\sqrt N)$ and $\ket s=\sin\theta_0\ket\tau+\cos\theta_0\ket{\tau_\perp}$. The first ideal updated state has angle $3\theta_0$, so
\begin{eqnarray}
D(\hat\rho_{\psi_1},\hat\rho_s)=\abs{\sin(2\theta_0)}=O(N^{-1/2}).
\end{eqnarray}
A learner returning the known state $\ket s$ without measuring any samples therefore meets any fixed trace-distance tolerance for sufficiently large $N$. Using this estimate for reflection does not yield the intended tripling trajectory. If the same estimate is used throughout $R=O(\log N)$ actual rounds, the dynamics is ordinary Grover rotation with angle $(2r+1)\theta_0$. Each actual state remains $O(R/\sqrt N)$ from $\ket s$, while the final marked-item probability is only $O(R^2/N)$. Thus, every round can satisfy a fixed learning-error tolerance and still fail to realize the desired speedup. This explicit example explains the need for Eq.~(\ref{eq:adaptive_error}).

\section{Classical reconstruction and the tomography benchmark}\label{app:tomography}\label{sec:SI_tomography}

The distinction between reconstructing a state and implementing an action is statistical as well as operational. A classical preparation description can be used repeatedly without new quantum samples. A consumed quantum program cannot. Here we state a precise reconstruction benchmark and explain how restricted state classes enter it.

\subsection{A self-contained average-fidelity bound} 

Let $d_M={{N+M-1} \choose {M}}$ be the dimension of the symmetric subspace of $M$ copies, with projector $\hat P_M$. Normalized Haar measure satisfies
\begin{eqnarray}
\int d\psi\,\hat\rho_\psi^{\otimes M}=\frac{\hat P_M}{d_M}.
\label{eq:SI_Haar_identity}
\end{eqnarray}
Any classical reconstruction scheme is represented by a POVM $\{\hat E_x\}$ on the copies and an estimated density operator $\hat\sigma_x$. Its Haar-averaged reconstruction fidelity is
\begin{eqnarray}
\overline F &=& \int d\psi\sum_x\tr{(\hat E_x\hat\rho_\psi^{\otimes M})}\tr{(\hat\sigma_x\hat\rho_\psi)} = \frac1{d_{M+1}}\sum_x\tr{[(\hat E_x\otimes\hat\sigma_x)\hat P_{M+1}]} \nonumber \\
	&\le&\frac1{d_{M+1}}\sum_x\tr{(\hat E_x\hat P_M)} = \frac{M+1}{M+N}.
\label{eq:SI_average_fidelity}
\end{eqnarray}
The inequality uses $\hat P_{M+1}\le\hat P_M\otimes\hat{\mathds{1}}$, positivity of $\hat E_x\otimes\hat\sigma_x$, $\tr{\hat\sigma_x}=1$, and POVM completeness. Integrals replace sums for continuous outcomes. The bound is attained by the covariant POVM with density $d_M\hat\rho_\phi^{\otimes M}d\phi$ on the symmetric support, followed by the estimate $\hat\rho_\phi$. Completeness follows from Eq.~(\ref{eq:SI_Haar_identity}); substitution gives equality in Eq.~(\ref{eq:SI_average_fidelity}). This is the familiar optimal pure-state estimation benchmark~\cite{hayashi2005reexamination}.

For pure estimates, $D(\hat\rho_\psi,\hat\rho_\phi)^2=1-\tr{(\hat\rho_\psi\hat\rho_\phi)}$. Hence, the mean squared trace error at most $\epsilon^2$ requires
\begin{eqnarray}
M\ge\frac{N-1}{\epsilon^2}-N.
\label{eq:SI_mean_squared_bound}
\end{eqnarray}
This statement uses a mean-risk criterion. High-probability trace-distance learning is a different criterion, for which the established minimax tomography results apply~\cite{haah2016sample,o2016efficient}. One must not convert a mean-risk identity into a confidence-dependent lower bound without an additional argument. Both criteria already require order-$N$ samples at fixed small error and fixed nontrivial confidence in the universal reconstruction problem. Requiring no-signaling cannot lower the optimum by enlarging the set of admissible quantum measurement strategies.

\subsection{Restricted circuit classes and the direction of minimax comparisons} 

For a specified circuit model, define the state class
\begin{eqnarray}
\mathcal S_{n,G}=\left\{\ket\psi=\hat U\ket0^{\otimes n}: \hat U\in\mathcal U_{n,G}\right\},
\label{eq:SI_circuit_class}
\end{eqnarray}
where $\mathcal U_{n,G}$ has at most $G$ two-qubit gates, with its allowed one-qubit gates, connectivity, and precision convention fixed. Covering and packing methods relate the sample complexity to the effective size of this class. Sharp bounded-gate results, including upper bounds and the qualifications needed for lower bounds, are established in Ref.~\cite{zhao2024learning}. They provide a computational benchmark rather than an independent consequence of causality.

For any trajectory family $\mathcal T\subseteq\mathcal S_{n,G}$, the minimax learning complexities satisfy
\begin{eqnarray}
M_\star(\mathcal T,\epsilon,\delta) \le M_\star(\mathcal S_{n,G},\epsilon,\delta).
\label{eq:SI_class_inclusion}
\end{eqnarray}
A learning algorithm that works for the larger class also works for its subset. Thus, a worst-case lower bound for the larger class does not imply that a particular trajectory, or each state along it, is hard. A separate packing within the actual promise class is required. Likewise, a lower bound on circuit size is not an upper bound on circuit size and cannot identify an optimal sample scaling by substitution into a learning lower bound.

For completeness, an information-theoretic route must keep both parts of its hypothesis explicit. If a collection of pure states is separated by order $\epsilon$ \emph{and} lies inside a ball of radius $O(\epsilon)$ around a pure reference state, its average has most of its weight on that reference. This can bound its entropy by an expression containing $h_2(O(\epsilon^2))+O(\epsilon^2)\log N$. Pairwise separation alone gives no such entropy upper bound: orthogonal states are a counterexample. Logarithmic dimension factors cannot be absorbed into universal constants. We therefore use the established minimax results, rather than an entropy estimate based only on separated states, for comparisons with universal or bounded-gate tomography.

\section{Source accounting and the oracle formulation}\label{app:oracles}\label{sec:SI_oracles}

Unstructured search encodes a unique marked computational basis state $\ket\tau$ in $\hat R_\tau=\hat{\mathds{1}}-2\ketbra\tau\tau$. Its resource model differs from a source that supplies arbitrary copies of an initially unknown state. To compare algorithms in a black-box query model, all oracle-dependent information used in training must be charged.

\begin{proposition}[Preparation-weighted learning budget]\label{prop:SI_query_budget}
Suppose an algorithm finds the marked item with fixed success probability at least $2/3$ for every oracle. Assume that its complete physical implementation, including training, state preparation, verification, and the final output, has a worst-case query upper bound
\begin{eqnarray}
Q_{\rm tot}\le Q_{\rm other}+\sum_{r=0}^{R-1}q_r m_r.
\label{eq:SI_query_upper}
\end{eqnarray}
Its initial state and classical advice must be independent of the marked item. Then
\begin{eqnarray}
Q_{\rm other}+\sum_r q_r m_r\ge c\sqrt N
\label{eq:SI_query_lower}
\end{eqnarray}
for a constant $c>0$ and sufficiently large $N$.
\end{proposition}
\begin{proof}
The standard quantum query lower bound for unstructured search gives $Q_{\rm tot}\ge c\sqrt N$~\cite{bennett1997strengths,zalka1999grover}. Combining it with the \emph{upper} bound in Eq.~(\ref{eq:SI_query_upper}) proves the result. Measurements, classical randomness, and adaptive controls are included in the standard model. The query upper bound and success premise must both be justified for the implementation being analysed.
\end{proof}

If $q_r\le q_{\max}=O(1)$, $Q_{\rm other}=O(R)$, and $R=\Theta(\log N)$, the proposition gives
\begin{eqnarray}
\sum_r m_r = \Omega(\sqrt N),
\quad
\frac{1}{R}\sum_r m_r = \Omega\left(\frac{\sqrt N}{\log N}\right).
\label{eq:SI_average_round}
\end{eqnarray}
This is a bound on the average allocated budget in the specified successful architecture, not on every round separately. If a copy requires more queries, its preparation weight $q_r$ must remain in Eq.~(\ref{eq:SI_query_lower}). If a source supplies oracle-dependent copies without charge, one has changed the oracle problem by adding a resource; the unassisted search lower bound cannot be applied to the enlarged input model without accounting for that resource.

Producing a new state from an already learned circuit may require no further oracle call once that description is available. The earlier queries that generated the description are still part of $Q_{\rm other}$ or a previous training budget. Conversely, when a fresh intermediate state requires another amplification step, the associated oracle call must be counted. Thus, the statement that a state depends on $\tau$ does not by itself determine a lower or upper query cost for each additional copy.

The universal coherent construction in this work uses $q=(k-1)/2$ calls to $\hat R_\Pi$. Setting $\hat\Pi=\ketbra\tau\tau$ and $k=\Theta(\sqrt N)$ gives $\Theta(\sqrt N)$ target-oracle calls, consistent with search optimality. Its initial-copy lower bound concerns a processor required to work for every unknown pure input. It cannot be applied as a lower bound on copies of the known uniform initialization alone. These distinctions are necessary to compare the two formulations without assigning the same meaning to different samples.

\subsection{Why a query lower bound is not a learned-gate lower bound} 

A short classical description can be expensive to discover and inexpensive to use. Once a marked bit string is known, a reflection about that basis state can be implemented by basis changes and a multi-controlled phase, with polynomial gate count in $n=\log_2N$. The cost of learning the string is not part of that circuit's gate count. More general uniform-plus-marked trajectory states also have a compact description once the string and angle are known. Consequently, query accounting alone does not force an exponential gate lower bound for the resulting learned reflections. General state-synthesis complexity is a separate problem with different quantifiers and must be stated for its actual state class.

\subsection{The scope of the causality--search connection} 

Bao, Bouland, and Jordan analyse specified modifications of quantum mechanics and relate their search power to their signaling capability~\cite{bao2016grover}. We use that work as physical context. The present causal obstruction instead follows directly from equal quantum input ensembles and remote steering. Our resource inequality uses the standard quantum search lower bound and a fully specified query implementation. No identification of circuit depth with a light-travel time is needed.

In particular, the identity $(\hat{\mathds{1}}\otimes\hat V)\ket\Phi=(\hat V^T\otimes\hat{\mathds{1}})\ket\Phi$ for a fixed maximally entangled vector does not let a remote party implement $\hat V^T$ on an arbitrary local input. The local marginal of that entangled resource is unchanged. The explicit steering construction used for the ensemble witness is different: it changes a decomposition of a fixed marginal, whose local distinguishability is precisely what the hypothetical nonlinear amplifier would introduce.

\section{Relation to simulation frameworks and logical scope}\label{app:scope}\label{sec:SI_context}

The operational distinction between a desired output state, a simulated channel, and a reusable classical description determines which established complexity bounds apply. Our task fixes the input to be copies of one unknown pure state and specifies one output depending on that same state. The registers are independent conditional on $\psi$, but their common unknown parameter supplies the program information. Accuracy is required for this prescribed state transformation; arbitrary independent data inputs are not part of the task.

Sample-based Hamiltonian simulation solves a stronger channel problem. Given copies of a generating state $\hat\rho$, it approximates the action of $e^{-it\hat\rho}$ on arbitrary data, including data entangled with a reference. Kimmel \textit{et al.}~\cite{kimmel2017hamiltonian} established the sample-complexity framework, and Go \textit{et al.}~\cite{go2025sample} provide a non-asymptotic analysis of density-matrix exponentiation and matching channel-simulation bounds in their specified regimes. Our upper bound uses a channel guarantee of this kind with $\hat\rho=\hat\rho_\psi$ and $t=\pi$. Its direct partial-swap proof controls errors on arbitrary intermediate registers. This sufficient construction does not make generic channel-simulation lower bounds necessary for our more restricted output task.

The samplizer framework of Wang and Zhang~\cite{wang2026samplizer} likewise converts reflection-query algorithms into sample-based simulations with controlled channel error. Their generic simulation theorem supplies the relevant context for the quadratic query-to-sample overhead. Their sample-to-query lifting framework~\cite{wang2025lifting} relates property-testing sample complexity to block-encoding query complexity. These results compare carefully specified access models; they do not identify copies with free calls to a preparation unitary. In particular, optimal accuracy dependence for a generic simulator cannot be transferred to a chosen state-dependent output without an additional reduction preserving its input promise and error criterion.

Our causal converse uses standard quantum state geometry, ordinary probabilistic mixing, a device defined on the full input space, and remote steering. It derives the ensemble inequality without assuming a CPTP processor in advance. Actual upper bounds are then supplied by CPTP constructions. No-signaling constrains which ensemble-sensitive learning prescriptions admit a physical completion; the quantum overlap law quantifies the resulting sample requirement. The general finite-scale theorem is an explicit consequence of this two-state inequality, rather than a new independent physical postulate.

The concrete analysis concerns the amplification task: shifted polygon ensembles yield an exact resolution threshold, their complete finite-copy distinguishability is evaluated, and a matching fixed-error copy scaling follows. On a known real great circle, frequency-pair filters further give matching dependence on amplification and accuracy. Every heralding record remains internal to an unconditional protocol. This restricted result does not close the accuracy gap for arbitrary complex pure inputs; impossibility of exact universal heralded amplification also supplies no inverse-accuracy lower bound by itself.

Finally, full tomography requests a classical description that can support subsequently selected operations. Consuming quantum samples to produce one target state need not reconstruct that description, so a dimension-independent control cost does not improve universal tomography bounds. A supplied preparation unitary and its inverse constitute another resource: they permit coherent operations unavailable from copies alone. If training states require calls to an unknown oracle, their preparation costs enter the total query budget. These distinctions express the connection between causal consistency and quantum control: the information required by a specified transformation must enter through a physically admissible, explicitly counted resource.

\subsection{Logical scope of the results}\label{sec:SI_scope} 

The causal threshold assumes the ordinary quantum state space, its density-operator description of unconditioned mixtures, remote ensemble preparation, and no-signaling. It compares preparations with exactly the same complete input state. The quantitative ensemble and two-state inequalities also admit a direct causal proof by completing unequal inputs into equal-state ensembles. That proof requires the device to be defined on arbitrary states of the full input registers, including the auxiliary states in the completion; accuracy is promised only on identical pure-state copies. CPTP contractivity provides an equivalent operational proof for quantum implementations. The upper construction uses explicit quantum operations. Thus, the optimal fixed-error copy scaling is established with quantum state-space assumptions; no claim is made that no-signaling alone selects this scaling in every possible physical theory.

All universal amplification statements concern pure inputs. For a general mixed density operator, $\hat{\mathds{1}}-2\hat\rho$ is not generally unitary, so it cannot be substituted for a Householder reflection in the same theorem. Sample-based simulation of $\exp(-it\hat\rho)$ remains meaningful, but it defines a different task. Similarly, the results concern deterministic output with a fixed available-copy budget. A heralded device must include both its success probability and its failure output; conditioning on success without charging its probability changes the operational problem.

The task-specific contribution is the explicit finite-copy ensemble analysis, the precision-optimal real-circle construction and its obstruction to exact universal heralded extension, and their connection to achievable amplification, adaptive accuracy allocation, and preparation-weighted learning. The finite-scale expansion theorem organizes the resulting necessary condition for more general state-dependent tasks. The general facts that no-signaling constrains ensemble-dependent operations, that trace distance contracts under quantum channels, that universal tomography has a dimension cost, and that reflection queries can be simulated from state samples are established ingredients. Their roles are separated so that the physical statement and the computational optimality claim can each be checked under its own assumptions.

\section*{References}


\end{document}